\documentclass[reprint,
 superscriptaddress,
 amsmath,amssymb,
 aps,
 pra,
 nofootinbib
]{revtex4-2}

\usepackage{overpic}
\usepackage{framed}
\usepackage{physics}
\clearpage{}\usepackage[uline]{hhtensor}
\usepackage{mathrsfs}
\usepackage{hyperref}
\usepackage{mathtools}
\usepackage{siunitx}
\usepackage{xcolor}
\usepackage{textgreek}
\newcommand*{\hbaraux}[2]{\sbox0{\mathsurround=0pt$#1\mkern-1mu\mathchar'26$}\mkern-1mu\lower.07\ht0\box0\mkern-8mu}
\renewcommand*{\hbar}{{\mathpalette\hbaraux\relax\mathrm{h}}}

\newcommand\Eq[1]{\text{Eq.~(\ref{#1})}}

\newcommand\FourierinvthreeD[1]{\mathscr{F}^{-1}_{3D}{\left\{#1\right\}}}

\newcommand\rr{\mathbf{r}}
\newcommand\rrbar{\mathbf{\bar{r}}}

\newcommand\pp{\mathbf{k}}

\newcommand\phat{\mathbf{\hat{\pp}}}

\newcommand\Fpplus{\mathbf{F}_+(\pp)}
\newcommand\Fpminus{\mathbf{F}_-(\pp)}
\newcommand\Gpplus{\mathbf{G}_+(\pp)}
\newcommand\Gpminus{\mathbf{G}_-(\pp)}
\newcommand\Gplambda{\mathbf{G}_{\lambda}(\pp)}
\newcommand\sphharm[3]{\mathrm{Y}_{#1#2}(#3)}

\newcommand\ff[1]{\mathrm{f}_{#1}(\pp)}
\newcommand\ffstar[1]{\mathrm{f}^*_{#1}(\pp)}
\newcommand\FF[1]{\mathrm{f}_{jm#1}(|\pp|)}
\newcommand\FFstar[1]{\mathrm{f}^*_{jm#1}(|\pp|)}

\newcommand\Flambdaykreg{\mathbf{F}^{\text{reg}}_{\lambda}(\yy,|\pp|)}
\newcommand\Flambdaykout{\mathbf{F}^{\text{out}}_{\lambda}(\yy,|\pp|)}
\newcommand\Flambdaykin{\mathbf{F}^{\text{in}}_{\lambda}(\yy,|\pp|)}
\newcommand\Glambdaykout{\mathbf{G}^{\text{out}}_{\lambda}(\yy,|\pp|)}
\newcommand\Glambdaykin{\mathbf{G}^{\text{in}}_{\lambda}(\yy,|\pp|)}

\newcommand\Flambdayk{\mathbf{F}_{\lambda}(\yy,|\pp|)}

\newcommand\Flambdaty{\mathbf{F}_{\lambda}(\yy,t)}

\newcommand\Glambdayk{\mathbf{G}_{\lambda}(\yy,|\pp|)}

\newcommand\Ayk{\mathbf{A}(\yy,|\pp|)}
\newcommand\Byk{\mathbf{B}(\yy,|\pp|)}
\newcommand\Ark{\mathbf{A}(\rr,|\pp|)}
\newcommand\Brk{\mathbf{B}(\rr,|\pp|)}

\newcommand\Frtplus{\mathbf{F}_+(\rr,t)}
\newcommand\Frtplusout{\mathbf{F}^{\text{out}}_+(\rr,t)}

\newcommand\Frtlambda{\mathbf{F}_{\lambda}(\rr,t)}

\newcommand\Frtminus{\mathbf{F}_-(\rr,t)}
\newcommand\Frtminusout{\mathbf{F}^{\text{out}}_-(\rr,t)}

\newcommand\Ert{\mathbf{E}(\rr,t)}

\newcommand\Brt{\mathbf{B}(\rr,t)}

\newcommand\Id{\mathrm{I}}
\newcommand\yy{\mathbf{y}}

\newcommand\intdSdD{\int_{\yy\in\partial D}\mathbf{\mathrm{d}S}(\yy)}

\newcommand\Frklambda{\mathbf{F}_\lambda(\rr,|\pp|)}
\newcommand\Grklambda{\mathbf{G}_\lambda(\rr,|\pp|)}
\newcommand\Fyklambda{\mathbf{F}_\lambda(\yy,|\pp|)}
\newcommand\Gyklambda{\mathbf{G}_\lambda(\yy,|\pp|)}

\newcommand\Frkhatklambda{\mathbf{F}_\lambda(|\rr|\phat,|\pp|)}
\newcommand\Grkhatklambda{\mathbf{G}_\lambda(|\rr|\phat,|\pp|)}
\newcommand\intdmodphbar{\int_{>0}^\infty \frac{\mathrm{d}|\pp|}{\hbar\cz|\pp|}\text{ }}
\newcommand\intdmodpenergy{\int_{>0}^\infty \mathrm{d}|\pp|\text{ }}
\newcommand\intdmodphelicity{\int_{>0}^\infty \frac{\mathrm{d}|\pp|}{\cz|\pp|}\text{ }}

\newcommand\intdpconfhbar{\int_{\mathbb{R}^3} \frac{\mathrm{d}^3 \pp}{\hbar\cz|\pp|}\text{ }}
\newcommand\intdpconfhbarwithoutzero{\int_{\mathbb{R}^3-\{\mathbf{0}\}} \frac{\mathrm{d}^3 \pp}{\hbar\cz|\pp|}\text{ }}

\newcommand\intdpnormwithoutzero{\int_{\mathbb{R}^3-\{\zerovec\}} \frac{\mathrm{d}^3 \pp}{\sqrt{(2\pi)^3}}\text{ }}

\newcommand\intdr{\int_{\mathbb{R}^3} {\mathrm{d}^3 \rr}\text{ }}
\newcommand\intdrbar{\int_{\mathbb{R}^3} {\mathrm{d}^3 \bar{\rr}}\text{ }}

\newcommand\zerovec{\mathbf{0}}

\newcommand\ii{\mathrm{i}}

\newcommand\muz{\text{\textmu}_{\text{0}}}
\newcommand\epsz{\text{\textepsilon}_{\text{0}}}
\newcommand\cz{\mathrm{c_0}}

\newcommand\Fplambda{\mathbf{F}_\lambda(\pp)}

\newcommand\rhat{\mathbf{\hat{r}}}

\newcommand\Fplambdarhat{\mathbf{F}_\lambda(|\pp|\rhat)}
\renewcommand\op[1]{\mathrm{#1}}
\newcommand{\mybraket}[2]{\langle#1|#2\rangle}
\newcommand{\mybraketout}[2]{{^{\text{out}}}\langle#1|#2\rangle{^{\text{out}}}}
\newcommand{\mybraketin}[2]{{^{\text{in}}}\langle#1|#2\rangle{^{\text{in}}}}
\newcommand{\mybraketreg}[2]{{^{\text{reg}}}\langle#1|#2\rangle{^{\text{reg}}}}
\newcommand{\sandwich}[3]{\langle#1|#2|#3\rangle}
\newcommand{\sandwichout}[3]{{^{\text{out}}}\langle#1|#2|#3\rangle^{\text{out}}}
\newcommand{\sandwichin}[3]{{^{\text{in}}}\langle#1|#2|#3\rangle^{\text{in}}}
\newcommand\brain[1]{{^{\text{in}}\hspace{-0.1cm}}\bra{#1}}
\newcommand\ketin[1]{\ket{#1}^{\text{in}}}
\newcommand\braout[1]{{^{\text{out}}\hspace{-0.1cm}}\bra{#1}}
\newcommand\brareg[1]{{^{\text{reg}}\hspace{-0.1cm}}\bra{#1}}
\newcommand\ketout[1]{\ket{#1}^{\text{out}}}
\newcommand\ketreg[1]{\ket{#1}^{\text{reg}}}
\newcommand{\eq}[1]{\begin{align}#1\end{align}}

\newcommand\jmax{j_{\text{max}}}

\newcommand\inout{\text{in/out}}
\clearpage{}
\usepackage{empheq}
\usepackage{comment}
\usepackage{booktabs}
\usepackage[capitalise]{cleveref}
\usepackage[customcolors]{hf-tikz}

\tikzset{offset definition 1/.style={
    above left offset={-0.4,0.6},
    below right offset={2.7,-0.5},
  },
   filling/.style={
    disable rounded corners=true,
    set fill color=white,
    set border color=black,
	line width=0.1mm,
  },
  box it 1/.style={
    offset definition 1,
    filling
  },
 }
\tikzset{offset definition 2/.style={
    above left offset={-0.1,0.4},
    below right offset={2.85,1.7},
  },
   filling/.style={
    disable rounded corners=true,
    set fill color=white,
    set border color=black,
	line width=0.1mm,
  },
  box it 2/.style={
    offset definition 2,
    filling
  },
 }

\begin{document}
\title{The electromagnetic scalar product in spatially-bounded domains}
\author{Maxim Vavilin}
\affiliation{Institut f\"ur Theoretische Festk\"orperphysik, Karlsruhe Institute of Technology, Kaiserstr. 12, 76131 Karlsruhe, Germany}
\author{Carsten Rockstuhl}
\affiliation{Institut f\"ur Theoretische Festk\"orperphysik, Karlsruhe Institute of Technology, Kaiserstr. 12, 76131 Karlsruhe, Germany}
\affiliation{Institute of Nanotechnology, Karlsruhe Institute of Technology, Kaiserstr. 12, 76131 Karlsruhe, Germany}
\author{Ivan Fernandez-Corbaton}
\email{ivan.fernandez-corbaton@kit.edu}
\affiliation{Institute of Nanotechnology, Karlsruhe Institute of Technology, Kaiserstr. 12, 76131 Karlsruhe, Germany}
\begin{abstract}
	Many physically interesting quantities of the electromagnetic field can be computed using the electromagnetic scalar product. However, none of the existing expressions for such scalar product are directly applicable when the fields are only known in a spatially-bounded domain, as is the case for many numerical Maxwell solvers. In here, we derive an expression for the electromagnetic scalar product between radiation fields that only involves integrals over closed spatial surfaces. The expression readily leads to formulas for the number of photons, energy, and helicity of generic polychromatic light pulses of incoming or outgoing character. The capabilities of popular Maxwell solvers in spatially-bounded computational domains are thereby augmented, for example, by a straightforward method for normalizing emitted fields so that they contain a single photon.
\end{abstract}
\maketitle
\section{Introduction and summary}
Quantum and classical nanophotonics benefit from advances in nanofabrication, such as the integration of molecules and quantum dots into complex systems, and the precise structuring of materials at the nanoscale \cite{Lodahl2015,Aharonovich2016,Hahn2019,Yang2021}. It is important that this progress is matched by theoretical advances enlarging the capabilities of computational tools such as numerical Maxwell solvers. These kind of solvers are often used for the design and optimization of functional devices that exploit modern fabrication possibilities. Generic numerical Maxwell solvers may be broadly classified as either time-domain or frequency-domain solvers. A very popular approach in time-domain solvers is the finite-difference time-domain method (FDTD) \cite{Taflove2005}, where Maxwell equations are discretized in space and time, and a time-marching algorithm is used to propagate the field from a source through a given photonic structure. Frequency-domain solvers, appropriate for linear problems, use the finite-element method (FEM) \cite{Monk2003} or the finite-difference frequency-domain method (FDFD) \cite{Rumpf2022}. Additionally, there are approaches adapted to specific settings. For example, plane wave expansion techniques are suitable for periodic structures \cite{Moharam1981}, and the boundary element method is particularly useful when considering localized scatterers \cite{Hohenester2012,Hohenester2022}. Another popular tool is the T-matrix formalism \cite{Waterman1965,Gouesbet2019,Mishchenko2020}, where, for a given object, a set of illuminations and their resulting scattered fields are used to build its T-matrix for later use. The T-matrix encodes the response of the object to generic illuminations. The upfront computational price is typically offset by the efficiency with which the response of a composite system can be obtained from the T-matrices of its components, in particular for periodic arrangements such as metasurfaces \cite{Gantzounis2006,Beutel2020}. Recently, the formalism has been reworked, and made to rest on an inherently polychromatic framework \cite{Vavilin2023}, as opposed to the typical monochromatic one. The polychromatic framework is directly applicable to the interaction of material objects with light pulses, and to the description of emissions from molecules or quantum dots. The T-matrices can then be seen as operators in $\mathbb{M}$, the Hilbert space of general polychromatic solutions of Maxwell equations.

A core element of the Hilbert space formalism is the electromagnetic scalar product \cite{Gross1964}, which has important uses beyond underpinning the mathematical structure of $\mathbb{M}$. For example, given an electromagnetic field $\ket{f}\in\mathbb{M}$, its norm squared $\mybraket{f}{f}$ is equal to the number of photons in the field \cite{Zeldovich1965}. The total energy and helicity contained in the field are equal to $\sandwich{f}{\op{H}}{f}$ and $\sandwich{f}{\Lambda}{f}$, where $\op{H}$ and $\Lambda$ are the energy and helicity operators, respectively. Actually, the total amount of any quantity represented by a self-adjoint operator $\Gamma$, such as energy, helicity, linear momentum, and angular momentum, is equal to $\sandwich{f}{\Gamma}{f}$. The scalar product can also be used to obtain the coefficient functions of the expansions of the electromagnetic fields in different basis, such as plane, spherical, or cylindrical waves.

Knowledge of the number of photons of a given field in a simulation can be used to re-scale such field so that it contains any desired number of photons, in particular one. This is needed in quantum nanophotonics to model single photon emitters, or the interaction of single photons with molecules, for example. Also, in a semi-classical or full-quantum treatment of the light-matter interaction \cite{Ge2015} using e.g., a Jaynes-Cummings model \cite{Waks2010}, one needs to know the number of photons contained in a modal field: The coupling strength of an emitter to a mode needs to be calculated from a normalized field distribution where the mode contains exactly one photon. In another example, the total number of photons and the helicity of the field radiated by an emitter nearby a (chiral) nanostructure is a quantity of interest in the study of luminescence enhancement, in particular in chiral luminescence enhancement \cite{Zambrana2016b}. While these physically interesting quantities can be computed with scalar products, the currently available expressions are not directly compatible with popular Maxwell solvers. None of the expressions is directly applicable to fields known only in a spatially-bounded domain, which is the information available when using for example COMSOL, MEEP, JCMsuite, CST, or Lumerical. Some of the expressions for the scalar product involve the fields in the entire infinitely extended space (\cite[Eq.~(1)]{Zeldovich1965},\cite[Eq.~(6)]{Gross1964}, \Eq{eq:rt}). Others require the coefficients of the expansion of the fields into a basis, such as plane waves or spherical waves, also known as multipolar fields (\cite{Moses1965b}, Eqs.~(\ref{eq:lmsp},\ref{eq:amsp}))
\begin{figure}[ht!]
        \includegraphics[width=0.8\linewidth]{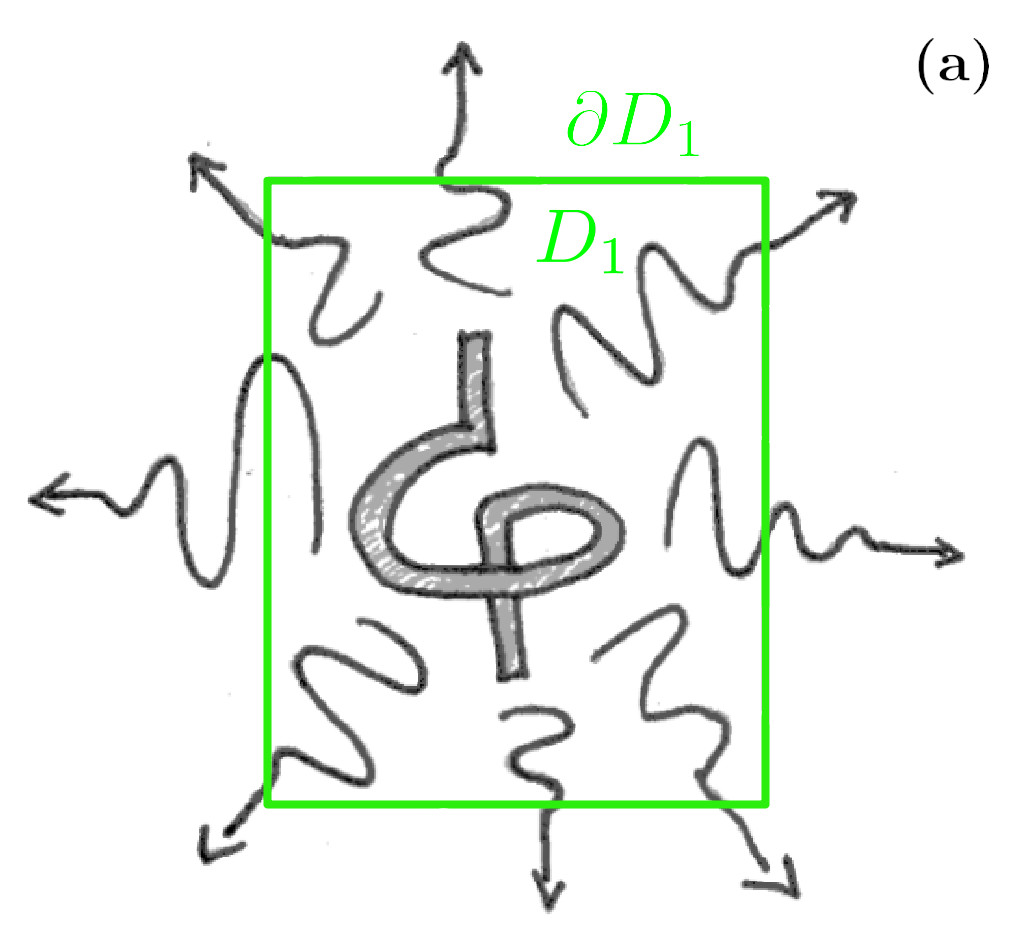}
        \includegraphics[width=0.65\linewidth,angle=-90]{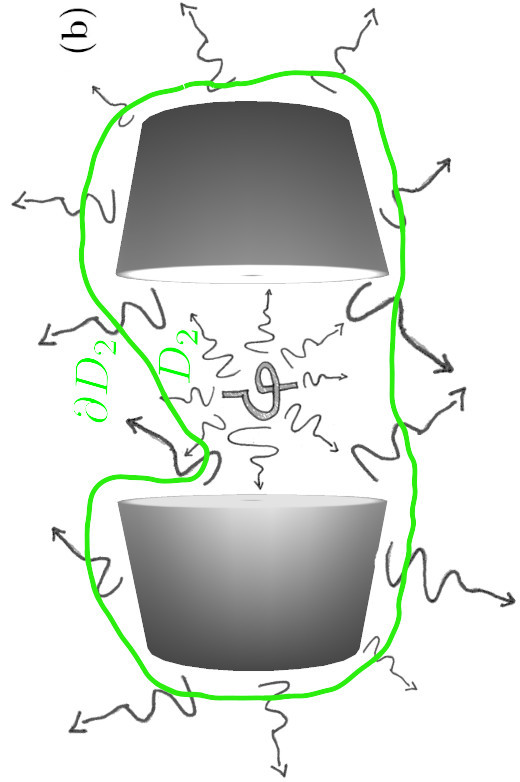}
	\caption{\label{fig:composition}In a numerical simulation, the electromagnetic scalar products resulting in the number of photons, energy, and helicity of the pulse emitted by the object in (a) can be obtained as integrals on the $\partial D_1$ surface using \cref{eq:formulaone,eq:formulatwo,eq:formulathree}. Adjustments to the simulated emission can then be done, for example, by scaling the field so that it contains a single photon. In (b), the adjusted emitter is placed close to other objects, which interact with the emission. Integrals on the $\partial D_2$ surface provide the same quantities for the total outgoing field.}
\end{figure}

To overcome such limitation, we derive here a new expression for the electromagnetic scalar product between two radiation fields, either incoming or outgoing. The new expression involves only the fields on a closed spatial surface (see Fig.~\ref{fig:composition}). For outgoing fields, the surface must enclose the sources of radiation, and for incoming fields, the surface must exclude the sources. The new expression readily leads to formulas for the number of photons, helicity, and energy, of a given radiation field. The expressions are given in SI units, and feature explicit physical constants, which facilitates their implementation. The formulas are derived for polychromatic fields with general frequency dependence, and are hence applicable to generic light pulses. The results for monochromatic simulations are the frequency-dependent densities of the different computed quantities, such as, for example, the density of photons per frequency.

The rest of the article is structured as follows. In Sec.~\ref{sec:surfint}, we derive the new expressions, which are numerically verified in Sec~\ref{sec:numver}. The new expression for the scalar product is written in \Eq{eq:result}, and used to derive formulas for the number of photons, helicity, and energy, in \Eq{eq:formulaone}, \Eq{eq:formulatwo}, and \Eq{eq:formulathree}, respectively. In Sec.~\ref{sec:practice}, we discuss practical aspects for using the formulas in the context of popular Maxwell solvers. Section~\ref{sec:conclusion} concludes the article. We expect the new expressions to be useful in classical and quantum computations in nanophotonics. 

The computer source codes used to produce the numerical results can be downloaded from \cite{SurfintCode2023}. 

\section{The scalar product between outgoing or incoming fields can be computed on a closed spatial surface\label{sec:surfint}}
SI units and the helical combinations of electric and magnetic fields for $\lambda=\pm$1 will be used throughout this article:
\begin{equation}
	\label{eq:lambda}
	\Frtlambda=\sqrt{\frac{\epsz}{2}}\left[\Ert+\ii\lambda \cz\Brt\right],
\end{equation}
where $\Ert$ and $\Brt$ are the total electric field and magnetic induction, respectively, and $\cz$ and $\epsz$ denote the speed of light and permittivity of vacuum, respectively. For the purposes of this work, a non-absorbing homogeneous and isotropic background medium different than vacuum can be accommodated in the formalism. One can just use the corresponding permittivity and permeability instead of the values for vacuum. Both $\Ert$ and $\Brt$ are complex fields, as explained below.

The helical $\Frtlambda$ fields can be built as the following sum of plane waves:
\begin{equation}
	\label{eq:frt}
		\Frtlambda=\intdpnormwithoutzero \Fplambda\exp(\ii\pp\cdot\rr-\ii\cz|\pp| t),
\end{equation}
where $\pp\cdot\Fplambda=0$, and, importantly, the time-harmonic angular frequency is restricted to positive values $\omega=\cz|\pp|>0$. The $\Frtlambda$ are eigenstates of the helicity operator with eigenvalue $+$1 and $-$1, and they split the electromagnetic field into its left-handed and right-handed circular polarization components, respectively. Such splitting works for general fields: Far fields, near fields, cavity modes, etc ... . The restriction to positive frequencies, which makes $\Ert$ and $\Brt$ necessarily complex valued, is crucial for $\Frtlambda$ to actually separate the two handedness of the fields. It readily follows from \Eq{eq:lambda} that, if $\Ert$ and $\Brt$ are real-valued, then $|\Frtplus|=|\Frtminus|$, which negates the handedness separation. The $\Frtlambda$ are the positive frequency restriction of the Riemann-Silberstein vectors \cite{Birula1996}, and their monochromatic components are also known as Beltrami fields \cite{Lakhtakia1994}.

Let us consider the conformally invariant scalar product for Maxwell fields \cite{Gross1964}, which can be written as:
\begin{equation}
	\label{eq:scalarproduct}
	\langle f|g\rangle=\intdpconfhbarwithoutzero \begin{bmatrix}\Fpplus\\\Fpminus\end{bmatrix}^\dagger \begin{bmatrix}\Gpplus\\\Gpminus\end{bmatrix},
\end{equation}
where $\hbar$ is the reduced Planck constant. By using the rules of Fourier transforms\footnote{In particular that from \cite[Eqs~(B.3,B.4)~and~Tab.~II~in~ I.B.2]{Cohen1997} it follows that $\FourierinvthreeD{\mathbf{F}_{\lambda}(\pp)\exp(-\ii\cz|\pp|t)\times\frac{1}{|\pp|}}=\frac{1}{2\pi^2}\intdrbar \mathbf{F}_{\lambda}(\bar{\rr},t)\times \frac{1}{|\rr-\rrbar|^2}$.}, \Eq{eq:scalarproduct} can also be written as:
\begin{equation}
	\label{eq:rt}
	\begin{split}
		&2\pi^2\hbar\cz\langle f|g\rangle=\\
		&\intdr\intdrbar \frac{\Frtplus^\dagger\mathbf{G}_+(\rrbar,t)+\Frtminus^\dagger\mathbf{G}_-(\rrbar,t)}{|\rr-\rrbar|^2}.
	\end{split}
\end{equation}

Other expressions of the scalar product can be written down using the coefficient functions of the expansions of the electromagnetic fields in different bases. For example
\begin{equation}
	\label{eq:lmsp}
	\langle f|g\rangle = \sum_{\lambda=\pm1} \int_{\mathbb{R}^3-\{\mathbf{0}\}} \frac{\text{d}^3 \pp}{|\pp|} \, \ffstar{\lambda}\mathrm{g}_\lambda (\pp),
\end{equation}
where the $\ff{\lambda}$ and $\mathrm{g}_\lambda (\pp)$ are the complex scalar coefficient functions of the plane wave expansions, and
\begin{equation}
	\label{eq:amsp}
	\langle f|g\rangle = \sum_{\lambda=\pm 1} \int_{>0}^{\infty} \text{d}|\pp|\, |\pp| \sum_{j=1}^{\infty} \sum_{m=-j}^j \, \FFstar{\lambda} \mathrm{g}_{jm\lambda}(|\pp|),
\end{equation}
where the $\FF{\lambda}$ and $\mathrm{g}_{jm\lambda}(|\pp|)$ are the coefficient functions of the expansions in spherical waves, also known as multipolar fields. The form of the expressions (\ref{eq:lmsp}) and (\ref{eq:amsp}) is achieved using the conventions in \cite{Vavilin2023}, which we include in App.~\ref{app:conventions}. 

The choice of scalar product is actually rather forceful. To begin with, this is the scalar product that produces the correct values for the fundamental quantities in electrodynamics. That is, for example, the result of $\sandwich{f}{\op{H}}{f}$ coincides with the result of the typical integral for the energy of the field $\intdr \left(\epsz|\mathbf{E}|^2+\frac{1}{\muz}|\mathbf{B}|^2\right)$, and the same is true for the other quantities \cite[Chap.~3,\S 9]{Birula1975}. Moreover, the meaningful frame-independent definition of the number of photons and of projective measurements is possible thanks to the conformal invariance of the chosen scalar product \cite[Sec.~3]{FerCor2022b}. For example, the number of photons $\mybraket{f}{f}$ is only meaningful and suitable for quantization if its value is the same under all the possible changes of reference frame allowed by Maxwell equations, that is, under all the transformations in the conformal group in 3+1 Minkowski space-time \cite{Bateman1910}. 

We highlight that none of the above expressions for the scalar product is directly applicable when the fields emitted by, or scattered off an object under a particular illumination, are available in numerical calculations in a spatially-bounded domain. An indirect method is possible by multipolar decomposition via surface integrals inside the domain \cite{Garcia2018}. However, this indirect route can become computationally expensive. One is first required to establish an appropriate truncation value for the multipolar order $j\le \jmax$ in \Eq{eq:amsp} by sequentially increasing $\jmax$ until some convergence criterium is fulfilled. The number of surface integrals is equal to $2{\jmax}^2+4\jmax$. The appropriate $\jmax$ often implies that hundreds of surface integrals need to be computed for each frequency, in particular for the incoming fields, and for fields scattered by wavelength-sized objects with moderate to high refractive index, where $\jmax$ will easily be higher than 10. Saving computational resources is particularly important in the context of optimization algorithms, where computations have to be done at each iteration because the object has changed. Moreover, in gradient-based optimizations, bypassing the multipolar decomposition can save one step in the derivative chain rule that connects the computed fields to the optimization target.

We will now obtain a new expression for the scalar product between two outgoing fields. Fields emitted from a quantum dot, and the fields scattered off a nanostructure under a given illumination are examples of outgoing fields. The new expression only involves the values of the fields on any piecewise smooth surface enclosing a compact volume containing the object, and is hence directly applicable to fields computed in a spatially-bounded domain (see Fig.~\ref{fig:composition}). The extension to incoming fields is straightforward and will be explained afterwards. For incoming fields, the surface must exclude the sources, and for outgoing fields, it must enclose the sources of radiation. 

We start by considering the asymptotic behavior of \Eq{eq:frt} for large $|\rr|$. To such end, we consider the expansion\footnote{Explicit definitions for the spherical Bessel functions $j_l(|\pp||\rr|)$, and for the spherical harmonics $\sphharm{l}{m}{\rhat}$ can be found in Eqs.~(3.111, 3.82) and Eq.~(3.53) of Ref.~\onlinecite{Jackson1998}, respectively.} of the $\rr$-dependent exponential in \Eq{eq:frt}:
\begin{equation}
	\label{eq:exp}
	\exp\left(\ii\pp\cdot\rr\right)=4\pi\sum_{l=0}^\infty\sum_{m=-l}^{l}\ii^l j_l(|\pp||\rr|)\sphharm{l}{m}{\phat}{\sphharm{l}{m}{\rhat}}^*.
\end{equation}

The large-$|\rr|$ behavior of \Eq{eq:exp} is only determined by the spherical Bessel function, as in \cite[Eq.~(9.89)]{Jackson1998}:
\begin{equation}
\label{eq:limit}
	\begin{split}
		&j_l(|\pp||\rr|)\eval_{|\rr|\rightarrow\infty} \rightarrow\frac{1}{|\pp||\rr|}\sin\left(|\pp||\rr|-\frac{l\pi}{2}\right)=\\
		&\frac{1}{|\pp||\rr|2\ii}\left[\exp\left(\ii|\pp||\rr|-\frac{\ii l\pi}{2}\right)-\exp\left(-\ii|\pp||\rr|+\frac{\ii l\pi}{2}\right)\right].
	\end{split}
\end{equation}
We take now only the outgoing part of the last line of \Eq{eq:limit}:
\begin{equation}
	(-\ii)^{l+1}\frac{\exp\left(\ii|\pp||\rr|\right)}{2|\pp||\rr|},
\end{equation}
and plug it into the right hand side of \Eq{eq:exp} in substitution of $j_l(|\pp||\rr|)$. After such substitution, we obtain:
\begin{equation}
	\label{eq:cool}
	\begin{split}
		&-\ii 2\pi\frac{\exp\left(\ii|\pp||\rr|\right)}{|\pp||\rr|} \sum_{l=0}^\infty\sum_{m=-l}^{l}\sphharm{l}{m}{\phat}{\sphharm{l}{m}{\rhat}}^*=\\
		&-\ii 2\pi\frac{\exp\left(\ii|\pp||\rr|\right)}{|\pp||\rr|} \delta(\cos\theta_\pp-\cos\theta_\rr)\delta(\phi_\pp - \phi_\rr),
	\end{split}
\end{equation}
	where $\phi_\pp(\phi_\rr)$ and $\theta_\pp(\theta_\rr)$ are the polar and azimuthal angles of $\pp$($\rr$) in spherical coordinates, respectively, and the equality follows from \cite[Eq.~(8.6-10)]{Tung1985}. The delta distributions clearly indicate that \Eq{eq:cool} cannot be used outside of integrals. In particular, it would not produce the correct result for a single component $\exp\left(\ii\mathbf{k_0}\cdot\rr\right)$. 

We can now take \Eq{eq:cool} into \Eq{eq:frt}, express the $\text{d}^3\pp$ volume integral in spherical coordinates\footnote{$\int \text{d}^3\pp =\int \text{d}|\pp||\pp|^2\int \text{d}\cos\theta\int \text{d}\phi=\int \text{d}|\pp||\pp|^2\int \text{d}\theta\sin\theta\int \text{d}\phi$.}, and manipulate it into:
{\small
\begin{equation}
	\label{eq:Frtlimitone}
	\begin{split}
		&\Frtlambda\eval_{|\rr|\rightarrow\infty} \rightarrow\\
		&\frac{-\ii}{\sqrt{2\pi}}\int_{|\pp|>0}^\infty \text{\hspace{-0.5cm}}d|\pp||\pp|^2\exp(-\ii\cz|\pp| t) \frac{\exp\left(\ii|\pp||\rr|\right)}{|\pp||\rr|}\int \text{d}\phat \Fplambda\delta(\phat - \rhat)\\
		&=\frac{-\ii}{\sqrt{2\pi}}\int_{|\pp|>0}^\infty \text{ }d|\pp||\pp|\exp(-\ii\cz|\pp| t) \frac{\exp\left(\ii|\pp||\rr|\right)}{|\rr|} \Fplambdarhat,
	\end{split}
\end{equation}
}
where $\int \mathrm{d}\phat$ stands for $\int_{-1}^1\text{ }\text{d}(\cos\theta_\pp)\int_{-\pi}^\pi \text{d}\phi_\pp$, and \mbox{$\delta(\phat - \rhat)$} for $\delta(\cos\theta_\pp-\cos\theta_\rr)\delta(\phi_\pp - \phi_\rr)$.

We now consider a different expression for $\Frtlambda$, namely its expansion into monochromatic fields $\Frklambda$,
\begin{equation}
	\label{eq:monochromcomp}
	\Frtlambda=\int_{>0}^\infty \frac{\text{d}|\pp|}{\sqrt{2\pi}}\text{ }\Frklambda \exp(-\ii\cz|\pp| t),
\end{equation}
and examine its asymptotic behavior:
\begin{equation}
	\label{eq:Frtlimittwo}
			\Frtlambda\eval_{|\rr|\rightarrow\infty}\hspace{-0.2cm}\rightarrow
			\int_{>0}^\infty\frac{\text{d}|\pp|}{\sqrt{2\pi}}\text{ } \exp(-\ii\cz|\pp| t) \Frklambda\eval_{|\rr|\rightarrow\infty}.
\end{equation}

The following result is readily obtained by comparing the last line of \Eq{eq:Frtlimitone} with \Eq{eq:Frtlimittwo}:
\begin{equation}
	-\ii |\pp| \frac{\exp\left(\ii|\pp||\rr|\right)}{|\rr|} \Fplambdarhat =  \Frklambda\eval_{|\rr|\rightarrow\infty}.
\end{equation}
Let us now use such result to work on the integrand in \Eq{eq:scalarproduct}:
\begin{equation}
	\begin{split}
		&\sum_{\lambda=\pm 1}\Fplambda^\dagger\Gplambda=\\
		&\sum_{\lambda=\pm 1}\left[\frac{|\rr|}{|\pp|} \Frkhatklambda\eval_{|\rr|\rightarrow\infty}\right]^\dagger\left[ \frac{|\rr|}{|\pp|} \Grkhatklambda\eval_{|\rr|\rightarrow\infty}\right]\\
		&=\sum_{\lambda=\pm 1}\left[\frac{|\rr|^2}{|\pp|^2} \Frkhatklambda^\dagger\Grkhatklambda\right]_{|\rr|\rightarrow\infty},
	\end{split}
\end{equation}
and now substitute it in \Eq{eq:scalarproduct}:
\begin{equation}
	\label{eq:almost}
	\begin{split}
		&\langle f|g\rangle=\sum_{\lambda=\pm 1}\intdpconfhbar\Fplambda^\dagger\Gplambda=\\
		&\sum_{\lambda=\pm 1}\intdpconfhbar\left[\frac{|\rr|^2}{|\pp|^2} \Frkhatklambda^\dagger\Grkhatklambda\right]_{|\rr|\rightarrow\infty}\\
		&=\sum_{\lambda=\pm 1}\intdmodphbar\\
		&\text{\hspace{1.8cm}}\left[\int \text{d}\phat |\rr|^2 \Frkhatklambda^\dagger\Grkhatklambda\right]_{|\rr|\rightarrow\infty},
	\end{split}
\end{equation}
where the third equality follows from splitting the $\text{d}^3\pp$ integral into its radial and angular parts. 

The expression inside the square brackets in the last line of \Eq{eq:almost} is an integral over the surface of a sphere with radius $|\rr|\rightarrow\infty$. The core result of this article is reached after using a result from \cite{Garcia2018} to exchange such integral\footnote{The following equations in \cite{Garcia2018} are missing a multiplication by -1 on their right hand sides: (21,23,24,25).} with the integral over any piecewise smooth surface $\partial D$ enclosing a compact volume containing the sources of radiation (see App.~\ref{app:ds}):
\begin{align}\tikzmarkin[box it 2]{a}
	\label{eq:result}
    &\langle f|g\rangle=\\ 
	&\hspace{-0.1cm}\sum_{\lambda=\pm 1}\hspace{-0.2cm}\left(-\tau\right)\ii\lambda\hspace{-0.1cm}\intdmodphbar \int_{\yy\in\partial D}\hspace{-0.7cm}\mathbf{\mathrm{d}S}(\yy)\cdot\left[\Flambdayk^*\times\Glambdayk\right]\nonumber
\tikzmarkend{a}\end{align}
The result in \Eq{eq:result} with $\tau=1$ is valid for outgoing fields. With $\tau=-1$ it is valid for incoming fields. The derivation for incoming fields is very similar to the one written for outgoing fields. It starts from the complex conjugate of \Eq{eq:frt}, and one takes then the incoming part of the plane wave in \Eq{eq:limit}, instead of the outgoing part. The end result is the same as \Eq{eq:result} but multiplied by -1. This change of sign comes from the sign difference between the inward and outward radiation conditions at infinity. For incoming fields the surface $\partial D$ must exclude the sources of incoming radiation.

Some clarifications are in order at this point. The scalar product applies to free fields, that is, fields that are not interacting with matter. A radiation field is free between the time when its sources have finished radiating and the time when it enters into contact with another material object. During such period, the result of \Eq{eq:rt} is actually independent of time. In practice, this means that for obtaining complete and undisturbed information in a time-domain simulation, the outgoing fields on the surface, $\mathbf{F}_{\lambda}(\yy,t)$ and $\mathbf{G}_{\lambda}(\yy,t)$, need to be recorded for as long as the emissions lasts, and that during such periods of time there should not be reflections of the emitted fields coming back into the volume enclosed by the surface. Similarly, for incoming fields, there should not be reflections going back out of the enclosed volume. It is precisely during the periods when a field is free that, in the {\em polychromatic} setting, the incoming and outgoing fields are equal to their regular versions \cite[Sec.~3.2.3]{Vavilin2023}. This follows from the fact that, in sharp contrast with their eternal monochromatic counterparts, incoming(outgoing) polychromatic fields are zero after(before) some instance of time. For free fields, one may as well use regular fields for computing scalar products between two incoming or two outgoing fields, and then it follows that \Eq{eq:amsp} and \Eq{eq:scalarproduct} can be used even though the expressions do not include evanescent plane waves. Regular fields are the sum of incoming and outgoing fields, and are often referred to as standing or stationary, on account of such fields not producing a net flux of energy or photons at spatial infinity.

We note that the derivation leading to \Eq{eq:result} cannot be adapted to include regular fields. The sum in \Eq{eq:cool} that leads to the angular delta function cannot be performed because of remaining $l$-dependence of $\ii^l \sin(|\pp||\rr|-l\pi/2)/(|\pp||\rr|)$. The derivation of \Eq{eq:result} from \Eq{eq:scalarproduct} requires that the fields are of pure outgoing character. Similarly, pure incoming character of the fields is required to obtain the corresponding expression of the scalar product as a surface integral. Actually, it is straightforward to show that if both $\Frtlambda$ and $\mathbf{G}_\lambda(\rr,t)$ are regular fields, the surface integral in \Eq{eq:result} vanishes (see App.~\ref{app:vanish}). This, of course, does not mean that the scalar product between any two regular fields is zero. It rather reflects the fact that such surface integrals cannot be used to compute the scalar product between two regular fields. 
The surface integral in \Eq{eq:result} also vanishes if one of the fields is outgoing and the other is incoming. Besides a direct derivation, this result can also be seen as follows. For a given helicity $\lambda$, and a point in the far field, the polarization vector of a plane wave whose momentum points from the origin to such point, is orthogonal to the polarization vector for a plane wave whose momentum points from such point towards the origin:
\begin{equation}
	\label{eq:inoutzero}
	\left[\mathbf{\hat{e}}_\lambda(\pp)\right]^\dagger\mathbf{\hat{e}}_{\lambda}(-\pp)=-\left[\mathbf{\hat{e}}_\lambda(\pp)\right]^\dagger\mathbf{\hat{e}}_{-\lambda}(\pp)=0,
\end{equation}
where the first equality follows from e.g., \cite[Eq.~(180)]{Vavilin2023}, and the second from the well-known orthogonality of the polarization vectors of two plane waves with the same momentum direction but opposite handedness.

The orthogonality between incoming and outgoing fields can be used to compute the scalar product between a given incoming(outgoing) first field, and a second field. The second field can be implemented as a regular field, in which case, \Eq{eq:result} will result in the scalar product between the first incoming(outgoing) field and, the incoming(outgoing) part of the second field. This possibility has been shown already for the particular case of monochromatic multipolar fields: The use of spherical Bessel functions instead of spherical Hankel functions was shown in \cite{Garcia2018} to produce the same result, and to be preferable because of a much improved numerical stability. 

Section~\ref{sec:numver} contains examples that illustrate and numerically verify these results. 

Using the new expression in \Eq{eq:result}, the number of photons, helicity, and energy of a given outgoing ($\tau=1$) or incoming ($\tau=-1$) field can be computed as:
\begin{widetext}
\begin{align}\tikzmarkin[box it 1]{a}
	\label{eq:formulaone}
		\text{Number of photons }\langle f|f\rangle=&\sum_{\lambda=\pm 1}(-\tau)\ii\lambda\intdmodphbar \int_{\yy\in\partial D}\hspace{-0.7cm}\mathbf{\mathrm{d}S}(\yy)\cdot\left[\Flambdayk^*\times\Flambdayk\right]\\
	\label{eq:formulatwo}
		\text{Helicity }\langle f|\Lambda|f\rangle=&\sum_{\lambda=\pm 1}(-\tau)\ii\intdmodphelicity \int_{\yy\in\partial D}\hspace{-0.7cm}\mathbf{\mathrm{d}S}(\yy)\cdot\left[\Flambdayk^*\times\Flambdayk\right]\\
	\label{eq:formulathree}
		\text{Energy }\langle f|\op{H}|f\rangle=&\sum_{\lambda=\pm 1}(-\tau)\ii\lambda\intdmodpenergy \int_{\yy\in\partial D}\hspace{-0.7cm}\mathbf{\mathrm{d}S}(\yy)\cdot\left[\Flambdayk^*\times\Flambdayk\right]
\tikzmarkend{a}\end{align}
\end{widetext}
where the second and third equalities follow from the respective actions of the helicity $\Lambda$, and energy $\op{H}$, operators on the rightmost set of monochromatic components $\Flambdayk$. Namely, helicity is just a multiplication by $\hbar\lambda$, and energy a multiplication by $\hbar\cz|\pp|$. The corresponding equation for momentum squared \cite{Lasa2023}, $\op{P}^2=\text{P}^2_x+\text{P}^2_y+\text{P}^2_z$, can readily be obtained from the action of the $\op{P}^2$ operator, which is a multiplication by $\hbar^2|\pp|^2$.

The equations for the energy and helicity actually contain the respective fluxes of each quantity. The real part of the term $(-\tau)\sum_{\lambda=\pm1} \ii \lambda \left[\Flambdayk^*\times\Flambdayk\right]$ in the energy formula in \Eq{eq:formulathree} can be shown to correspond to the Poynting vector, and the real part of the term $(-\tau)\sum_{\lambda=\pm1} \ii \left[\Flambdayk^*\times\Flambdayk\right]$ in the helicity formula in \Eq{eq:formulatwo} can be shown to correspond to the helicity flux vector \cite[Sec.~1(a)]{Vesperinas2017}. The imaginary parts must vanish in both cases because the average values $\langle f|\op{H}|f\rangle$, and $\langle f|\Lambda|f\rangle$ must be real numbers, as both $\op{H}$ and $\Lambda$ are self-adjoint with respect to this scalar product.
\begin{widetext}

\begin{figure}[h!]
        \includegraphics[width=\linewidth]{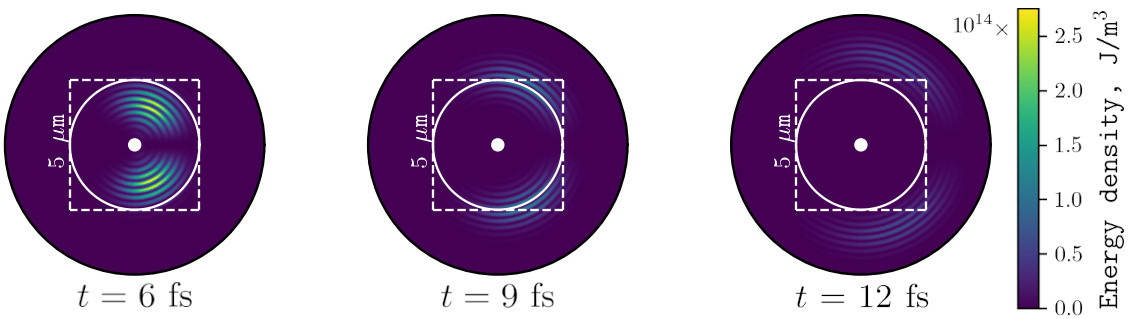}
	\caption{\label{fig:pulse}Energy density of the outgoing pulse from \Eq{eq:fielddef} at different times, plotted in the $zx$-plane with horizontal $z$-axis and vertical $x$-axis. The number of photons, helicity, and energy of the pulse can be computed with electromagnetic scalar products implemented as surface integrals on generic surfaces. The same results are obtained on spherical and cubical surfaces. The sphere centered at the origin has a radius of $2.5~\si{\mu\m}$ and is drawn with a solid line. The cube has sides of length $5.0~\si{\mu\m}$ and is drawn with dashed lines.}
\end{figure}
\begin{table}[h!]
{\small
\begingroup
\setlength{\tabcolsep}{10pt} \renewcommand{\arraystretch}{1.5} \begin{tabular}{l*{6}{c}r}
& $\mybraketout{f}{f}$  & $\sandwichout{f}{\Lambda}{f}$, \si{\joule\second}  & $\sandwichout{f}{H}{f}$, \si{\joule}  \\
\hline
Reference & $2.7841638840385884 \times 10^{16} $ & $-9.787001828407123 \times 10^{-19}$  & $0.011633883766510636$ \\
Sphere & $2.7841638841872064\times 10^{16}$& $-9.787001829974468 \times 10^{-19}$ & $0.011633883767253363$  \\
Sphere, shifted in $x$ & $2.7841598095399108\times 10^{16}$& $-9.78697157861713 \times 10^{-19}$ & $0.011633864885618727$  \\
Sphere, shifted in $y$ & $2.78415980953991\times 10^{16}$& $-9.786971578617121 \times 10^{-19}$ & $0.011633864885618714$  \\
Sphere, shifted in $z$ & $2.784163884061041\times 10^{16}$& $-9.78700182864386 \times 10^{-19}$ & $0.011633883766622862$  \\
Cube & $2.779822499549024 \times 10^{16}$ & $-9.76576139048995 \times 10^{-19}$ & $0.011615046402458412$
\end{tabular}
\endgroup
}
\caption{\label{table:compare}Number of photons, helicity and energy computed in six different ways. A reference value computed in a conventional way, and \cref{eq:formulaone,eq:formulatwo,eq:formulathree} computed with integrals on different closed surfaces: A sphere centered at the origin with radius $2.5~\si{\mu\m}$, spheres of the same radius but displaced in the positive $x-$, $y-$ and $z-$directions by $1.5~\si{\mu\m}$, and the surface of a cube. The units of each quantity are written in the first row, after the comma.}
\end{table}
\begin{table}[h!]
\begingroup
\setlength{\tabcolsep}{10pt} \renewcommand{\arraystretch}{1.5} \begin{tabular}{l*{6}{c}r}
& ${}^\text{out}\mybraket{f}{g}^{\text{out}}$ & ${}^\text{out}\mybraket{f}{g}^{\text{in}}$  & ${}^\text{out}\mybraket{f}{g}^{\text{reg}}$ \\
\hline
	Sphere  & $5.842862654438759\times 10^{15}$ & $-1.2349056843014804 \times 10^{-2}$ & $5.842862654438761 \times 10^{15}$ \\ 
	Sphere, shifted in $x$   & $5.842858857885211 \times 10^{15}$ & $5.8384755088009945 \times 10^{-3}$ & $5.842858857885217 \times 10^{15}$ \\ 
	Sphere, shifted in $y$   & $5.842858857885208 \times 10^{15}$ & $5.734217017572405 \times 10^{-3}$ & $5.842858857885215 \times 10^{15}$ \\ 
	Sphere, shifted in $z$   & $5.842862654438791 \times 10^{15}$ & $13.899064370402524$ & $5.842862654438798 \times 10^{15}$ \\ 
	Cube & $5.835536831748690 \times 10^{15}$ & $1.9192729636550432$  & $5.835536831748691 \times 10^{15}$ 
\end{tabular}
\endgroup
	\caption{\label{table:compare2}Scalar product between an outgoing field $\ketout{f}$ and the regular, incoming, and outgoing versions of a different field.}
\end{table}
\end{widetext}

\section{Numerical verification\label{sec:numver}}
For the rest of the article, we extend our notation to indicate explicitly the regular $\ketreg{f}$, incoming $\ketin{f}$, or outgoing $\ketout{f}$ character of electromagnetic fields. The relation $\ketreg{f}=\ketin{f}+\ketout{f}$ holds in our conventions (App.~\ref{app:conventions} and \cite{Vavilin2023}). The character of a given field is easily appreciated in its decomposition into multipolar fields. Such fields feature spherical Bessel functions for regular fields, and spherical Hankel functions of the first or second kind for outgoing or incoming fields, respectively. Substituting $j_l(\cdot)$ by $h_l^1(\cdot)/2$ changes the version of a field from regular to outgoing; substituting $h_l^1(\cdot)/2$ by $h_l^2(\cdot)/2$, changes it from outgoing to incoming, and so on. The same holds for cylindrical vector wave functions, which feature Bessel and Hankel functions. 
We first numerically verify \cref{eq:formulaone,eq:formulatwo,eq:formulathree} by computing the quantities contained in a given outgoing electromagnetic field. Consider an outgoing electromagnetic pulse defined by its two helicity components $\ketout{f}\equiv\left\{\Frtplusout,\Frtminusout\right\}$:
\begin{equation}
	\begin{split}
&\Frtplusout =\\
		&A\sqrt{\epsz} \int_{>0}^\infty \text{d}|\pp| \, |\pp| \, \exp\left(-\frac{(|\pp|-k_1)^2}{2\Delta^2}\right) \mathbf{S}^\text{out}_{331}(|\pp|,\rr,t),\\
		&\Frtminusout=\\
		&A\sqrt{\epsz} \int_{>0}^\infty \text{d}|\pp|\, |\pp| \, \exp\left(-\frac{(|\pp|-k_2)^2}{2\Delta^2}\right)\mathbf{S}^\text{out}_{2-2-1}(|\pp|,\rr,t), \label{eq:fielddef}
	\end{split}
\end{equation}
with a constant of $A=4\times10^{10}$~\si{\nano\meter}, characteristic pulse time span $\Delta^{-1} = 2~\si\fs$, and center wavelengths $\frac{2\pi}{k_1} = 800~\si{\nm}$ and $\frac{2\pi}{k_2} = 400~\si{\nm}$. For visualization purposes, we assume that the outgoing field is generated by sources confined inside a sphere of radius $200~\si{nm}$, which can be seen in the center of the plots in Fig.~\ref{fig:pulse}. The explicit definition of $\mathbf{S}^\text{out}_{jm\lambda}(|\pp|,\rr,t)$, which are multipolar fields and include the harmonic time dependence $\exp\left(-\ii\cz|\pp|t\right)$, can be found in App.~\ref{app:conventions}. Figure~\ref{fig:pulse} shows the energy density of the outgoing pulse from \Eq{eq:fielddef} at three different times.

We compute the amount of photons, helicity, and energy contained in the field by using \cref{eq:formulaone,eq:formulatwo,eq:formulathree} on five different surfaces: A spherical surface $\partial D_1$ of diameter $5.0~\si{\mu\m}$, the surface of a cube $\partial D_2$ with sides of $5.0~\si{\mu\m}$, both centered in the origin of the reference frame, as seen in Fig.~\ref{fig:pulse}, and then the three spherical surfaces that result from shifting the centered one by $1.5~\si{\mu\m}$ in the positive x-, y- and z-directions.

The integration over the spheres is implemented as a Riemann sum with discretization of polar and azimuthal angles in $N_\theta = 400$ and $N_\phi = 200$ equidistant points, respectively. The integration over the surface of the cube is performed as a Riemann sum with $N_x=N_y=N_z =200$ points. All integrals over $|\pp|$ are computed as a Riemann sum with $N_k = 200$ equidistant discretization points in the region $6.6~\si{\mu m}^{-1}\leq |\pp| \leq 17.0~\si{\mu m}^{-1}$, which covers the most significant part of the spectrum of the field.

Table~\ref{table:compare} shows the comparison of the results of \cref{eq:formulaone,eq:formulatwo,eq:formulathree} compared to the result of a conventional method based on \Eq{eq:amsp}, which we use as reference (see App.~\ref{app:reference}). The agreement is excellent, with numerical noise affecting the third significant digit. The closer agreement between the reference and the integrals on the spherical surfaces is likely because the reference is computed with the coefficients of the expansion of the fields in multipolar fields, which fit naturally to spherical surfaces. Importantly, additional calculations show that the results are identical, up to the same levels of numerical noise, when the lengths that determine the integration surfaces of the cube and the centered sphere are increased ten times, or decreased 20 times. In the latter case, the integration surfaces are very close to the \SI{200}{\nano\meter} spherical volume in the center of Fig.~\ref{fig:pulse}. 

We now numerically confirm other useful results. Namely, that the incoming and outgoing field types are orthogonal, and that, as a consequence, the value of the scalar product between a given outgoing field with the regular and outgoing versions of another given field are equal ${}^\text{out}\mybraket{f}{g}^{\text{reg}} = {}^\text{out}\mybraket{f}{g}^{\text{out}}$. We will use the previously defined $\ketout{f}$ in \Eq{eq:fielddef}, and the regular, incoming and outgoing versions of the following field $\ket{g}^{\text{reg/in/out}}\equiv\left\{\mathbf{G}^{\text{reg/in/out}}_+(\rr, t),\mathbf{G}^{\text{reg/in/out}}_-(\rr, t)\right\}$:
\begin{equation}
\begin{split}
	&\mathbf{G}^{\text{reg/in/out}}_+(\rr, t) =\\
	&A\sqrt{\epsz} \int_{>0}^\infty \text{d}|\pp| \, |\pp|\, \exp\left(-\frac{(|\pp|-k_3)^2}{2\Delta^2}\right) \mathbf{S}^\text{reg/in/out}_{331}(|\pp|,\rr,t), \\
	&\mathbf{G}^{\text{reg/in/out}}_-(\rr,t) =\zerovec,
\end{split}
	\end{equation}
with central wavelength $\frac{2 \pi}{k_3} =$ \SI{600}{\nano\meter}. 

Table~\ref{table:compare2} contains the numerical results which again show an excellent agreement with the expectations. We have similarly verified that $\brain{f}g\rangle^{\text{reg}}=\brain{f}g\rangle^{\text{in}}$.

\section{Practical considerations\label{sec:practice}}
In time-domain simulations such as FDTD, the complex-valued electric and magnetic fields are often available as time and spatially-dependent vector functions. This is the case in MEEP \cite{Oskooi2010}, for example. If only real-valued fields are available, their complex versions can be obtained by combining the results of two separate simulations. Here, we shall write the incident field as a product between a time-harmonic carrier field and an envelope whose Fourier transform expresses the launched spectrum. Then, the complex field can be obtained by running the simulations twice and ensuring that the carrier fields oscillate with a $\pi/2$ phase offset with respect to each other in these two simulations. Practically, once a cosine dependency is chosen and once a sine dependency. Then, summing the fields resulting from the cosine simulation with $\ii$ times the fields resulting from the sine simulation produces the desired complex fields. The $\Frtlambda$ can then be obtained from \Eq{eq:lambda}, and the $\Flambdayk$ fields on the surface $\partial D$ can be obtained by inverting \Eq{eq:monochromcomp}:
\begin{equation}
	\Flambdayk=\int_{-\infty}^{\infty}\frac{\text{d}t}{\sqrt{2\pi}}\Flambdaty\exp\left(\ii\cz|\pp|t\right),
\end{equation}
where the integration limits will be finite in practice because the fields on the surface will only be non-zero during a bounded period of time. Similarly, the limits in the $\text{d}|\pp|$ integrals in \cref{eq:formulaone,eq:formulatwo,eq:formulathree}, from $>0$ to $\infty$, will in practice be reduced to the finite bandwidth of the simulated fields. The formalism in Sec.~\ref{sec:surfint} is then directly applicable to numerical techniques such as FDTD, where the fields have a non-zero bandwith. However, numerical simulations that consider a single frequency of the field, such as FEM, are also commonly used. In the latter case, the formalism in Sec.~\ref{sec:surfint} can be used and interpreted as follows. In \cref{eq:result,eq:formulaone,eq:formulatwo,eq:formulathree}, we may move the $\int\text{d}|\pp|$ to the left, up to immediately after the equal sign. Then, the integrand under the $\int\text{d}|\pp|$ can be interpreted as the density of the corresponding quantity per wavenumber. The value of such densities at a discrete frequency is what can be obtained from a monochromatic simulation, where the $\mathbf{F}_\lambda(\yy,|\pp|)$ are easily accessible.

The formulas in \cref{eq:formulaone,eq:formulatwo,eq:formulathree} apply in particular to radiation from emitters such as molecules, quantum dots, or artificial meta-atoms excited by an external illumination, embedded in photonic systems. In such case, the value of the fields on a surface enclosing the emitter and the surrounding structures of interest are readily available in simulations (see Fig.~\ref{fig:composition}). The case of radiation from antennas fed by electric currents is treated similarly, and $\partial D$ can be conveniently chosen as the surface of the volume that defines the antenna, which must already be defined in the simulation domain. The quantities obtained with the formulas can be of interest by themselves, and also constitute a necessary step for further theoretical treatments. For example, when dealing with quantum-optical systems, the field from an emitter may need to be normalized in amplitude so that it contains exactly a single photon. Such field normalization is an initial step e.g., in the Jaynes-Cummings formalism to study strong coupling in nanophotonic systems \cite{Waks2010}, or in other modal approaches \cite{Ge2015}. In current approaches, the value of an integral over the whole infinitely extended space involving the field profile identified as a mode needs to be known. This value is frequently approximated by integrating across a finite volume while changing its size, and extrapolating the result of the integral to the infinite volume. However, this is notoriously prone to errors, especially for systems with large radiative losses such as plasmonic antennas. With our approach, the required normalization can be readily performed using the result of \Eq{eq:formulaone}. Similarly, the energy radiated by a given modal field, which is part of the definition of the mode volume, is equal to surface integral in \Eq{eq:formulatwo}, which avoids considering another integral in an infinitely large volume. 

When the illumination source is outside the computation domain, we speak of a scattering scenario, where a scattered outgoing field is produced by the interaction of the specified illumination with the object under study. Quantization of light in scattering scenarios is also of interest \cite{Savasta2002,Oppermann2018}. Formulas in \cref{eq:formulaone,eq:formulatwo,eq:formulathree} are directly applicable to the scattered field, which is available in simulations as the total field outside the object minus the illumination.

Scattering scenarios are also amenable to the study of transfer of fundamental quantities between light and matter. The amount of fundamental quantities such as energy is typically different in the incoming and outgoing fields, and the difference can potentially be transferred onto the object. The change of any quantity represented by a self-adjoint operator $\Gamma$ can be readily be written as:
\begin{equation}
	\label{eq:delta}
		\Delta\Gamma=\brain{f}\Gamma\ketin{f}-\braout{g}\Gamma\ketout{g},
\end{equation}
where $\ketin{f}$ is the incoming field and $\ketin{g}$ the outgoing field, which, with the aid of the scattering operator $\op{S}$ and transfer operator $\op{T}$, or T-matrix, can be written as\footnote{The difference with the typical convention $\op{S}=\Id+2\op{T}$ is due to the conventions in \cite{Vavilin2023}, which are more appropriate for the polychromatic case. See also App.~\ref{app:conventions}.}: 
\begin{equation}
	\label{eq:gout}
	\begin{split}
		\ketout{g}&=\op{S}\ketin{f}=\left(\Id+\op{T}\right)\ketin{f}=\ketout{f}+\op{T}\ketin{f}\\&=\ketout{f}+\ketout{h},
	\end{split}
\end{equation}
where $\ketout{h}=\op{T}\ketin{f}$ is the scattered field. Substitution of the last line of \Eq{eq:gout} in \Eq{eq:delta} results in:
\begin{equation}
	\label{eq:dgammascat}
	\begin{split}
		&\Delta\Gamma=\brain{f}\Gamma\ketin{f}-\\
		&\left[\sandwichout{f}{\Gamma}{f}+2\mathbb{R}\left\{\sandwichout{f}{\Gamma}{h}\right\}+\sandwichout{h}{\Gamma}{h}\right]\\
		&=-2\mathbb{R}\left\{\sandwichout{f}{\Gamma}{h}\right\}-\sandwichout{h}{\Gamma}{h},
	\end{split}
\end{equation}
where one uses that $\Gamma$ is self-adjoint, and that $\brain{f}\Gamma\ketin{f}=\braout{f}\Gamma\ketout{f}$, as deduced from \cite[Sec.~2.3.2]{Vavilin2023}. When $\Gamma$ represents the identity, the energy, or the helicity operators, the last line of \Eq{eq:dgammascat} could potentially be computed with \cref{eq:formulaone,eq:formulatwo,eq:formulathree} because it only involves outgoing fields. There is, however, an obstacle. While the value of the scattered field $\ketout{h}$ on a surface is explicitly available from numerical simulations in bounded domains, the outgoing version of the illumination $\ketout{f}$ is typically not. The illumination is specified as a regular field, often called incident field, without singularities inside the computational domain. This is in sharp contrast to the case of emitters within the simulation domain which, when modeled as point-like emitters, do contain singularities that are numerically avoided by excluding a small volume around such points. The obstacle is overcome by the fact that the surface integral in \Eq{eq:result} between an outgoing field and a regular field is equal to the scalar product between the outgoing field and the outgoing part of the regular field. Such result has been numerically verified in Sec.~\ref{sec:numver}. Therefore, the explicitly available regular incident field can be used to compute $\sandwichout{f}{\Gamma}{h}$ as:
\begin{equation}
	\label{eq:outreg}
\sandwichout{f}{\Gamma}{h}=\brareg{f}\Gamma\ketout{h}.
\end{equation}
For example, when $\Gamma$ is the identity we have that:
{\small
\begin{equation}
	\label{eq:outregsp}
	\begin{split}
		&\mybraketout{f}{h}=\\
		&\sum_{\lambda=\pm 1}-\ii\lambda\intdmodphbar \int_{\yy\in\partial D}\hspace{-0.7cm}\mathbf{\mathrm{d}S}(\yy)\cdot\left[\Flambdaykreg^*\times\mathbf{H}^\text{out}_\lambda(\yy,|\pp|)\right],
	\end{split}
\end{equation}
}
where the $\Flambdaykreg$ are the {\em regular} incident fields. Continuing with the example of the number of photons, we can then use \Eq{eq:outreg} and \Eq{eq:outregsp} to write:
{\small
\begin{equation}
	\begin{split}
		&\mybraketin{f}{f}-\mybraketout{g}{g}=\\
		&2\mathbb{R}\left\{\sum_{\lambda=\pm 1}\ii\lambda\intdmodphbar \int_{\yy\in\partial D}\hspace{-0.7cm}\mathbf{\mathrm{d}S}(\yy)\cdot\left[\Flambdaykreg^*\times\mathbf{H}^\text{out}_\lambda(\yy,|\pp|)\right]\right\}\\
		&+\left\{\sum_{\lambda=\pm 1}\ii\lambda\intdmodphbar \int_{\yy\in\partial D}\hspace{-0.7cm}\mathbf{\mathrm{d}S}(\yy)\cdot\left[\mathbf{H}^\text{out}_\lambda(\yy,|\pp|)^*\times\mathbf{H}^\text{out}_\lambda(\yy,|\pp|)\right]\right\}.
	\end{split}
\end{equation}
}
The formulas for the helicity and the energy changes are similarly obtained, and read:
{\small
\begin{equation}
	\begin{split}
		&\sandwichin{f}{\Lambda}{f}-\sandwichout{g}{\Lambda}{g}=\\
		&2\mathbb{R}\left\{\sum_{\lambda=\pm 1}\ii\intdmodphelicity\int_{\yy\in\partial D}\hspace{-0.7cm}\mathbf{\mathrm{d}S}(\yy)\cdot\left[\Flambdaykreg^*\times\mathbf{H}^\text{out}_\lambda(\yy,|\pp|)\right]\right\}\\
		&+\left\{\sum_{\lambda=\pm 1}\ii\intdmodphelicity\int_{\yy\in\partial D}\hspace{-0.7cm}\mathbf{\mathrm{d}S}(\yy)\cdot\left[\mathbf{H}^\text{out}_\lambda(\yy,|\pp|)^*\times\mathbf{H}^\text{out}_\lambda(\yy,|\pp|)\right]\right\},
	\end{split}
\end{equation}
}
and
{\small
\begin{equation}
	\begin{split}
		&\sandwichin{f}{\op{H}}{f}-\sandwichout{g}{\op{H}}{g}=\\
		&2\mathbb{R}\left\{\sum_{\lambda=\pm 1}\ii\lambda\intdmodpenergy \int_{\yy\in\partial D}\hspace{-0.7cm}\mathbf{\mathrm{d}S}(\yy)\cdot\left[\Flambdaykreg^*\times\mathbf{H}^\text{out}_\lambda(\yy,|\pp|)\right]\right\}\\
		&+\left\{\sum_{\lambda=\pm 1}\ii\lambda\intdmodpenergy \int_{\yy\in\partial D}\hspace{-0.7cm}\mathbf{\mathrm{d}S}(\yy)\cdot\left[\mathbf{H}^\text{out}_\lambda(\yy,|\pp|)^*\times\mathbf{H}^\text{out}_\lambda(\yy,|\pp|)\right]\right\},
	\end{split}
\end{equation}
}
respectively.

\section{Conclusion\label{sec:conclusion}}
This work uses the abstract formalism of the electromagnetic Hilbert space to augment the capabilities of numerical solvers of Maxwell equations in spatially-bounded domains. A new expression of the electromagnetic scalar product that involves only the fields on a closed spatial boundary allows the computation of the number of photons, helicity, and energy of incoming or outgoing radiation fields with general time-dependence. We expect the new expressions to be useful for classical and quantum computations in nanophotonics. For example, the calculation of the number of photons in numerically obtained fields allows one to re-scale such fields so that they contain a single photon. In another example, the total number of photons and the helicity of the field radiated by an emitter nearby a (chiral) nanostructure can be readily used to quantify luminescence enhancement, in particular chiral luminescence enhancement.

The computer source codes used to produce the numerical results can be downloaded from \cite{SurfintCode2023}. 

\begin{acknowledgments}
	The authors warmly thank Dr.~Xavier Garcia-Santiago for numerous useful discussions about the subject matter of this article, and for reading it and providing valuable feedback. This work was funded by the Deutsche Forschungsgemeinschaft (DFG, German Research Foundation) -- Project-ID 258734477 -- SFB 1173, and by the Helmholtz Association via the Helmholtz program ``Materials Systems Engineering'' (MSE).
\end{acknowledgments}

\appendix
\section{Conventions\label{app:conventions}}
This appendix contains the conventions for field expansions that we use in this article. They are taken from \cite{Vavilin2023}. 

The electric field is expanded into plane waves of well-defined helicity $\ket{\pp \,\lambda}$ as:
\begin{equation}
	\label{eq:xpans}
	\Ert=\sum_{\lambda=\pm1 }\int \frac{\text{d}^3 \pp}{|\pp|}\, \ff{\lambda} \, \ket{\pp \,\lambda},
\end{equation}
and the plane waves are defined as:
\begin{equation}
	\begin{split}
		&\ket{\pp \,\lambda}\equiv\\
		&\sqrt{\frac{\cz\hbar}{ \epsilon_0}}\, \frac{1}{\sqrt{2}} \frac{1}{\sqrt{(2\pi)^3}}\, |\pp| \, \mathbf{\hat{e}}_\lambda({\phat}) \exp(- \ii |\pp|\cz t ) \exp(i \pp \cdot \rr).
	\end{split}
\end{equation}
We highlight the factor of $|\pp|$ in the definition of the plane waves, which ensures that they transform unitarily under Lorentz transformations, and the factor of $1/|\pp|$ in \Eq{eq:xpans}, which makes the volume measure $\frac{\text{d}^3 \pp}{|\pp|}$ invariant under transformations in the Poincar\'e group.

The expansion in multipoles of well-defined helicity reads:
\begin{equation}
	\begin{split}
		&{\Ert}^{\text{reg/in/out}} \equiv\\
		&\int_0^\infty \text{d}|\pp| \, |\pp| \, \sum_{\lambda=\pm 1} \sum_{j=1}^{\infty} \sum_{m=-j}^j \, \FF{\lambda} \, \ket{|\pp| j m \lambda}^{\text{reg/in/out}},
	\end{split}
\end{equation}
and the regular, incoming, and outgoing multipoles $\ket{|\pp| j m \lambda}^{\text{reg/in/out}}$ are defined as:

\begin{widetext}
	\begin{equation}
		\begin{split}
		\label{eq:mpdef}
			&\ket{|\pp|jm\lambda}^{\text{reg}}\equiv\mathbf{S}^\text{reg}_{jm\lambda}(|\pp|,\rr,t)=\\
			&- \sqrt{\frac{\cz\hbar}{\epsilon_0}} \frac{1}{\sqrt{2\pi}} \, |\pp| \, \ii^j  \times\Big(  \exp(-\ii |\pp| \cz t)\, \mathbf{N}^{\text{reg}}_{jm}(|\pp||\rr|, \hat{\rr}) + \lambda \,\exp(-\ii |\pp|\cz t) \,  \mathbf{M}^{\text{reg}}_{jm}(|\pp||\rr|, \rhat ) \Big),\\
			&\ket{|\pp|jm\lambda}^{\text{in/out}}\equiv\mathbf{S}^\text{in/out}_{jm\lambda}(|\pp|,\rr,t)=\\
			&- \frac{1}{2} \sqrt{\frac{\cz\hbar}{\epsilon_0}} \frac{1}{\sqrt{2\pi}} \, |\pp| \, \ii^j\times \Big(  \exp(-\ii |\pp| \cz t)\, \mathbf{N}^{\inout}_{jm}(|\pp||\rr|, \hat{\rr}) + \lambda \,\exp(-\ii |\pp|\cz t) \,  \mathbf{M}^{\inout}_{jm}(|\pp||\rr|, \rhat ) \Big),
		\end{split}
	\end{equation}
\end{widetext}
where the $\mathbf{M}$ and $\mathbf{N}$ have the usual definitions (see e.g. \cite[Eqs.~(50,51)]{Vavilin2023}). We note the extra factor of $1/2$ in the definition of the incoming and outgoing multipoles, with respect to the regular multipoles. With this conventions, the decomposition of a regular field into its incoming and outgoing components holds in the following way: $\ketreg{f}=\ketin{f}+\ketout{f}$. The typical relation between the T-matrix and the S-matrix changes then from $\op{S}=\op{I}+2\op{T}$ to $\op{S}=\op{I}+\op{T}$.

\section{Detailed steps\label{app:ds}}
We now write the detailed steps leading from \Eq{eq:almost} to \Eq{eq:result}. In Ref.~\onlinecite{Garcia2018}, it was shown that, given two monochromatic electric fields $\Ayk$ and $\Byk$, both meeting the outgoing radiation conditions at infinity, the following holds:
\begin{equation}
	\label{eq:myto}
	\begin{split}
	&-2\ii|\pp|\lim_{|\rr|\rightarrow\infty}\int \text{d}\rhat |\rr|^2 \Brk^\dagger\Ark=\\
		&\int_{\yy\in\partial D}\hspace{-0.7cm}\mathbf{\mathrm{d}S}(\yy)\cdot\left\{\left[\nabla\times \Ayk\right]\times\Byk^*\right.\\
		&\left.-\left[\nabla\times \Byk\right]^*\times \Ayk\right\},
	\end{split}
\end{equation}
which is \cite[Eq.~(21)]{Garcia2018} in slightly different notation, except for a difference in sign, which is incorrect in \cite{Garcia2018}. The following equations in \cite{Garcia2018} are missing a multiplication by -1 on their right hand sides: (21,23,24,25). Equation~(\ref{eq:myto}) states that an integral over the surface of a sphere, when the radius tends to infinity, coincides with a different integral over a general surface $\partial D$ that encloses the radiating matter. The surface should be piecewise smooth and enclose a compact volume.

To reach the form in \Eq{eq:result} from \Eq{eq:almost}, we first consider that if $\Ark$ and $\Brk$ are fields of well-defined but opposite circular polarization handedness, their far fields are point-wise orthogonal\footnote{At each point of the far field, $\Ark$ and $\Brk$ will be essentially determined by a single plane wave with the same wavevector but opposite polarization handedness, hence the point-wise orthogonality.}, and hence the left hand side of \Eq{eq:myto} vanishes, and its right hand side must consequently be zero. This means that \Eq{eq:myto} must hold separately for each helicity, and we can write:
\begin{equation}
	\label{eq:fff}
	\begin{split}
	&-2\ii|\pp|\lim_{|\rr|\rightarrow\infty}\int \text{d}\rhat |\rr|^2 \Frklambda^\dagger\Grklambda=\\
		&\int_{\yy\in\partial D}\hspace{-0.7cm}\mathbf{\mathrm{d}S}(\yy)\cdot\left\{\left[\nabla\times \Gyklambda\right]\times\Fyklambda^*\right.\\
		&-\left.\left[\nabla\times \Fyklambda\right]^*\times \Gyklambda\right\}.
	\end{split}
\end{equation}
We now divide each side by $|\pp|$ and use that $\frac{\nabla\times}{|\pp|}$ is the helicity operator for monochromatic fields, which implies that $\frac{\nabla\times}{|\pp|}\Fyklambda=\lambda \Fyklambda$, and $\frac{\nabla\times}{|\pp|}\Gyklambda=\lambda \Gyklambda$, we can then re-write \Eq{eq:fff} as:
\begin{equation}
	\begin{split}
		&\lim_{|\rr|\rightarrow\infty}\int \text{d}\rhat |\rr|^2 \Frklambda^\dagger\Grklambda = \\
		&-\ii\lambda\intdSdD\cdot\left[\Flambdayk^*\times\Glambdayk\right],  
	\end{split}
\end{equation}
which is the substitution that takes us from the last line of \Eq{eq:almost} to \Eq{eq:result}.

\section{Reference values for scalar products\label{app:reference}}
With the definitions in \Eq{eq:mpdef}, the $\ketout{f}$ in Eq.~(\ref{eq:fielddef}) can be written as:
\begin{equation}
	\begin{split}
		\ketout{f} = A\sqrt{\epsz}\int_0^\infty &\text{d}|\pp| \, |\pp| \,\Big[ \exp(-\frac{(|\pp|-k_1)^2}{2\Delta^2}) \ket{|\pp|\,331}+\\
		& \exp\left(-\frac{(|\pp|-k_2)^2}{2\Delta^2}\right) \ket{|\pp|\, 2-2-1} \Big].
	\end{split}
\end{equation}

Since multipoles are orthogonal unless their discrete $jm\lambda$ labels coincide, the scalar products computed with \Eq{eq:amsp} reduce to: \eq{
	&\mybraketout{f}{f} = \sum_{\lambda=\pm 1} \sum_{j=1}^{\infty} \sum_{m=-j}^j \int^\infty_0 \text{d}|\pp|\, |\pp|\, \abs{\FF{\lambda}}^2= \nonumber \\
& A^2\epsz\int^\infty_0 \text{d}|\pp|\, |\pp|\,\,\Big[ \exp\left(-\frac{(|\pp|-k_1)^2}{\Delta^2}\right) + \exp\left(-\frac{(|\pp|-k_2)^2}{\Delta^2}\right) \Big],
}
for the number of of photons,
\eq{
	&\sandwichout{f}{\Lambda}{f} = \hbar \sum_{\lambda=\pm 1} \lambda \sum_{j=1}^{\infty} \sum_{m=-j}^j \int^\infty_0 \text{d}|\pp|\, |\pp|\, \abs{\FF{\lambda}}^2= \nonumber \\
 & A^2\epsz\hbar \int^\infty_0 \text{d}|\pp|\, |\pp|\, \abs{\FF{\lambda}}^2 \,\Big[ \exp\left(-\frac{(|\pp|-k_1)^2}{\Delta^2}\right) \nonumber\\ &- \exp\left(-\frac{(|\pp|-k_2)^2}{\Delta^2}\right) \Big],
 }
for the helicity, and
 \eq{
	 \sandwichout{f}{\op{H}}{f} =\hbar \cz \sum_{\lambda=\pm 1} \sum_{j=1}^{\infty}& \sum_{m=-j}^j \int^\infty_0 \text{d}|\pp|\, |\pp|^2\, \abs{\FF{\lambda}}^2 \nonumber \\
= A^2\epsz\hbar \cz \int^\infty_0 \text{d}|\pp|\, |\pp|^2\,\,&\Big[ \exp\left(-\frac{(|\pp|-k_1)^2}{\Delta^2}\right) \nonumber\\ &+ \exp\left(-\frac{(|\pp|-k_2)^2}{\Delta^2}\right) \Big]}
for the energy.

\section{Vanishing of the surface integrals for regular fields\label{app:vanish}}
Let us try to use \Eq{eq:result} to compute the scalar product between two regular fields: $\mybraketreg{f}{g}$. The first problem is that one cannot really choose an appropriate value of $\tau$. If we ignore this first difficulty, and insist on using the surface integrals with regular fields, we can use that a regular field is the sum of its outgoing and incoming versions to obtain the following integrands in \Eq{eq:result}:
\begin{equation}
		\left[\Flambdaykout+\Flambdaykin\right]^*\times\left[\Glambdaykout+\Glambdaykin\right].
\end{equation}
For the final result, however, only the following terms contribute:
\begin{equation}
		{\Flambdaykout}^*\times\Glambdaykout+{\Flambdaykin}^*\times\Glambdaykin,
\end{equation}
because, as argued in the paragraph containing \Eq{eq:inoutzero}, the terms that feature the cross product between an incoming field and an outgoing field will not contribute to the overall result. Then, for any value of $\tau$, the overall integral expression will actually be proportional to:
\begin{equation}
\mybraketout{f}{g}-\mybraketin{f}{g}
\end{equation}
, which is equal to zero because $\mybraketout{f}{g}=\mybraketin{f}{g}$ (\cite[Sec.~2.3.2]{Vavilin2023}). The vanishing of the result has been verified numerically.


\begin{thebibliography}{37}\makeatletter
\providecommand \@ifxundefined [1]{\@ifx{#1\undefined}
}\providecommand \@ifnum [1]{\ifnum #1\expandafter \@firstoftwo
 \else \expandafter \@secondoftwo
 \fi
}\providecommand \@ifx [1]{\ifx #1\expandafter \@firstoftwo
 \else \expandafter \@secondoftwo
 \fi
}\providecommand \natexlab [1]{#1}\providecommand \enquote  [1]{``#1''}\providecommand \bibnamefont  [1]{#1}\providecommand \bibfnamefont [1]{#1}\providecommand \citenamefont [1]{#1}\providecommand \href@noop [0]{\@secondoftwo}\providecommand \href [0]{\begingroup \@sanitize@url \@href}\providecommand \@href[1]{\@@startlink{#1}\@@href}\providecommand \@@href[1]{\endgroup#1\@@endlink}\providecommand \@sanitize@url [0]{\catcode `\\12\catcode `\$12\catcode
  `\&12\catcode `\#12\catcode `\^12\catcode `\_12\catcode `\%12\relax}\providecommand \@@startlink[1]{}\providecommand \@@endlink[0]{}\providecommand \url  [0]{\begingroup\@sanitize@url \@url }\providecommand \@url [1]{\endgroup\@href {#1}{\urlprefix }}\providecommand \urlprefix  [0]{URL }\providecommand \Eprint [0]{\href }\providecommand \doibase [0]{https://doi.org/}\providecommand \selectlanguage [0]{\@gobble}\providecommand \bibinfo  [0]{\@secondoftwo}\providecommand \bibfield  [0]{\@secondoftwo}\providecommand \translation [1]{[#1]}\providecommand \BibitemOpen [0]{}\providecommand \bibitemStop [0]{}\providecommand \bibitemNoStop [0]{.\EOS\space}\providecommand \EOS [0]{\spacefactor3000\relax}\providecommand \BibitemShut  [1]{\csname bibitem#1\endcsname}\let\auto@bib@innerbib\@empty
\bibitem [{\citenamefont {Lodahl}\ \emph {et~al.}(2015)\citenamefont {Lodahl},
  \citenamefont {Mahmoodian},\ and\ \citenamefont {Stobbe}}]{Lodahl2015}\BibitemOpen
  \bibfield  {author} {\bibinfo {author} {\bibfnamefont {P.}~\bibnamefont
  {Lodahl}}, \bibinfo {author} {\bibfnamefont {S.}~\bibnamefont {Mahmoodian}},\
  and\ \bibinfo {author} {\bibfnamefont {S.}~\bibnamefont {Stobbe}},\
  }\bibfield  {title} {\bibinfo {title} {Interfacing single photons and single
  quantum dots with photonic nanostructures},\ }\href
  {https://doi.org/10.1103/RevModPhys.87.347} {\bibfield  {journal} {\bibinfo
  {journal} {Rev. Mod. Phys.}\ }\textbf {\bibinfo {volume} {87}},\ \bibinfo
  {pages} {347} (\bibinfo {year} {2015})}\BibitemShut {NoStop}\bibitem [{\citenamefont {Aharonovich}\ \emph {et~al.}(2016)\citenamefont
  {Aharonovich}, \citenamefont {Englund},\ and\ \citenamefont
  {Toth}}]{Aharonovich2016}\BibitemOpen
  \bibfield  {author} {\bibinfo {author} {\bibfnamefont {I.}~\bibnamefont
  {Aharonovich}}, \bibinfo {author} {\bibfnamefont {D.}~\bibnamefont
  {Englund}},\ and\ \bibinfo {author} {\bibfnamefont {M.}~\bibnamefont
  {Toth}},\ }\bibfield  {title} {\bibinfo {title} {Solid-state single-photon
  emitters},\ }\href {https://doi.org/10.1038/nphoton.2016.186} {\bibfield
  {journal} {\bibinfo  {journal} {Nature Photonics}\ }\textbf {\bibinfo
  {volume} {10}},\ \bibinfo {pages} {631} (\bibinfo {year} {2016})}\BibitemShut
  {NoStop}\bibitem [{\citenamefont {Hahn}\ \emph {et~al.}(2019)\citenamefont {Hahn},
  \citenamefont {Mayer}, \citenamefont {Thiel},\ and\ \citenamefont
  {Wegener}}]{Hahn2019}\BibitemOpen
  \bibfield  {author} {\bibinfo {author} {\bibfnamefont {V.}~\bibnamefont
  {Hahn}}, \bibinfo {author} {\bibfnamefont {F.}~\bibnamefont {Mayer}},
  \bibinfo {author} {\bibfnamefont {M.}~\bibnamefont {Thiel}},\ and\ \bibinfo
  {author} {\bibfnamefont {M.}~\bibnamefont {Wegener}},\ }\bibfield  {title}
  {\bibinfo {title} {3-d laser nanoprinting},\ }\href@noop {} {\bibfield
  {journal} {\bibinfo  {journal} {\textit{Opt. Photon. News}}\ }\textbf
  {\bibinfo {volume} {30}},\ \bibinfo {pages} {28} (\bibinfo {year}
  {2019})}\BibitemShut {NoStop}\bibitem [{\citenamefont {Yang}\ \emph {et~al.}(2021)\citenamefont {Yang},
  \citenamefont {Mayer}, \citenamefont {Bunz}, \citenamefont {Blasco},\ and\
  \citenamefont {Wegener}}]{Yang2021}\BibitemOpen
  \bibfield  {author} {\bibinfo {author} {\bibfnamefont {L.}~\bibnamefont
  {Yang}}, \bibinfo {author} {\bibfnamefont {F.}~\bibnamefont {Mayer}},
  \bibinfo {author} {\bibfnamefont {U.~H.~F.}\ \bibnamefont {Bunz}}, \bibinfo
  {author} {\bibfnamefont {E.}~\bibnamefont {Blasco}},\ and\ \bibinfo {author}
  {\bibfnamefont {M.}~\bibnamefont {Wegener}},\ }\bibfield  {title} {\bibinfo
  {title} {Multi-material multi-photon 3d laser micro- and nanoprinting},\
  }\href@noop {} {\bibfield  {journal} {\bibinfo  {journal} {\textit{Light:
  Advanced Manufacturing}}\ }\textbf {\bibinfo {volume} {2}},\ \bibinfo {pages}
  {296} (\bibinfo {year} {2021})}\BibitemShut {NoStop}\bibitem [{\citenamefont {Taflove}\ \emph {et~al.}(2005)\citenamefont
  {Taflove}, \citenamefont {Hagness},\ and\ \citenamefont
  {Piket-May}}]{Taflove2005}\BibitemOpen
  \bibfield  {author} {\bibinfo {author} {\bibfnamefont {A.}~\bibnamefont
  {Taflove}}, \bibinfo {author} {\bibfnamefont {S.~C.}\ \bibnamefont
  {Hagness}},\ and\ \bibinfo {author} {\bibfnamefont {M.}~\bibnamefont
  {Piket-May}},\ }\bibfield  {title} {\bibinfo {title} {Computational
  electromagnetics: the finite-difference time-domain method},\ }\href@noop {}
  {\bibfield  {journal} {\bibinfo  {journal} {The Electrical Engineering
  Handbook}\ }\textbf {\bibinfo {volume} {3}},\ \bibinfo {pages} {15} (\bibinfo
  {year} {2005})}\BibitemShut {NoStop}\bibitem [{\citenamefont {Monk}(2003)}]{Monk2003}\BibitemOpen
  \bibfield  {author} {\bibinfo {author} {\bibfnamefont {P.}~\bibnamefont
  {Monk}},\ }\href@noop {} {\emph {\bibinfo {title} {Finite element methods for
  Maxwell's equations}}}\ (\bibinfo  {publisher} {Oxford University Press},\
  \bibinfo {year} {2003})\BibitemShut {NoStop}\bibitem [{\citenamefont {Rumpf}(2022)}]{Rumpf2022}\BibitemOpen
  \bibfield  {author} {\bibinfo {author} {\bibfnamefont {R.~C.}\ \bibnamefont
  {Rumpf}},\ }\href@noop {} {\emph {\bibinfo {title} {Electromagnetic and
  Photonic Simulation for the Beginner: Finite-Difference Frequency-Domain in
  MATLAB{\textregistered}}}}\ (\bibinfo  {publisher} {Artech House},\ \bibinfo
  {year} {2022})\BibitemShut {NoStop}\bibitem [{\citenamefont {Moharam}\ and\ \citenamefont
  {Gaylord}(1981)}]{Moharam1981}\BibitemOpen
  \bibfield  {author} {\bibinfo {author} {\bibfnamefont {M.~G.}\ \bibnamefont
  {Moharam}}\ and\ \bibinfo {author} {\bibfnamefont {T.~K.}\ \bibnamefont
  {Gaylord}},\ }\bibfield  {title} {\bibinfo {title} {Rigorous coupled-wave
  analysis of planar-grating diffraction},\ }\href
  {https://doi.org/10.1364/JOSA.71.000811} {\bibfield  {journal} {\bibinfo
  {journal} {J. Opt. Soc. Am.}\ }\textbf {\bibinfo {volume} {71}},\ \bibinfo
  {pages} {811} (\bibinfo {year} {1981})}\BibitemShut {NoStop}\bibitem [{\citenamefont {Hohenester}\ and\ \citenamefont
  {Trügler}(2012)}]{Hohenester2012}\BibitemOpen
  \bibfield  {author} {\bibinfo {author} {\bibfnamefont {U.}~\bibnamefont
  {Hohenester}}\ and\ \bibinfo {author} {\bibfnamefont {A.}~\bibnamefont
  {Trügler}},\ }\bibfield  {title} {\bibinfo {title} {Mnpbem – a matlab
  toolbox for the simulation of plasmonic nanoparticles},\ }\href
  {https://doi.org/https://doi.org/10.1016/j.cpc.2011.09.009} {\bibfield
  {journal} {\bibinfo  {journal} {Computer Physics Communications}\ }\textbf
  {\bibinfo {volume} {183}},\ \bibinfo {pages} {370} (\bibinfo {year}
  {2012})}\BibitemShut {NoStop}\bibitem [{\citenamefont {Hohenester}\ \emph {et~al.}(2022)\citenamefont
  {Hohenester}, \citenamefont {Reichelt},\ and\ \citenamefont
  {Unger}}]{Hohenester2022}\BibitemOpen
  \bibfield  {author} {\bibinfo {author} {\bibfnamefont {U.}~\bibnamefont
  {Hohenester}}, \bibinfo {author} {\bibfnamefont {N.}~\bibnamefont
  {Reichelt}},\ and\ \bibinfo {author} {\bibfnamefont {G.}~\bibnamefont
  {Unger}},\ }\bibfield  {title} {\bibinfo {title} {Nanophotonic resonance
  modes with the nanobem toolbox},\ }\href
  {https://doi.org/https://doi.org/10.1016/j.cpc.2022.108337} {\bibfield
  {journal} {\bibinfo  {journal} {Computer Physics Communications}\ }\textbf
  {\bibinfo {volume} {276}},\ \bibinfo {pages} {108337} (\bibinfo {year}
  {2022})}\BibitemShut {NoStop}\bibitem [{\citenamefont {Waterman}(1965)}]{Waterman1965}\BibitemOpen
  \bibfield  {author} {\bibinfo {author} {\bibfnamefont {P.~C.}\ \bibnamefont
  {Waterman}},\ }\bibfield  {title} {\bibinfo {title} {Matrix formulation of
  electromagnetic scattering},\ }\href {https://doi.org/10.1109/PROC.1965.4058}
  {\bibfield  {journal} {\bibinfo  {journal} {Proc. IEEE}\ }\textbf {\bibinfo
  {volume} {53}},\ \bibinfo {pages} {805} (\bibinfo {year} {1965})}\BibitemShut
  {NoStop}\bibitem [{\citenamefont {Gouesbet}(2019)}]{Gouesbet2019}\BibitemOpen
  \bibfield  {author} {\bibinfo {author} {\bibfnamefont {G.}~\bibnamefont
  {Gouesbet}},\ }\bibfield  {title} {\bibinfo {title} {T-matrix methods for
  electromagnetic structured beams: A commented reference database for the
  period 2014–2018},\ }\href
  {https://doi.org/https://doi.org/10.1016/j.jqsrt.2019.04.004} {\bibfield
  {journal} {\bibinfo  {journal} {Journal of Quantitative Spectroscopy and
  Radiative Transfer}\ }\textbf {\bibinfo {volume} {230}},\ \bibinfo {pages}
  {247} (\bibinfo {year} {2019})}\BibitemShut {NoStop}\bibitem [{\citenamefont {Mishchenko}(2020)}]{Mishchenko2020}\BibitemOpen
  \bibfield  {author} {\bibinfo {author} {\bibfnamefont {M.~I.}\ \bibnamefont
  {Mishchenko}},\ }\bibfield  {title} {\bibinfo {title} {Comprehensive thematic
  t-matrix reference database: a 2017-2019 update},\ }\href
  {https://doi.org/https://doi.org/10.1016/j.jqsrt.2019.106692} {\bibfield
  {journal} {\bibinfo  {journal} {Journal of Quantitative Spectroscopy and
  Radiative Transfer}\ }\textbf {\bibinfo {volume} {242}},\ \bibinfo {pages}
  {106692} (\bibinfo {year} {2020})}\BibitemShut {NoStop}\bibitem [{\citenamefont {Gantzounis}\ and\ \citenamefont
  {Stefanou}(2006)}]{Gantzounis2006}\BibitemOpen
  \bibfield  {author} {\bibinfo {author} {\bibfnamefont {G.}~\bibnamefont
  {Gantzounis}}\ and\ \bibinfo {author} {\bibfnamefont {N.}~\bibnamefont
  {Stefanou}},\ }\bibfield  {title} {\bibinfo {title}
  {Layer-multiple-scattering method for photonic crystals of nonspherical
  particles},\ }\href {https://doi.org/10.1103/PhysRevB.73.035115} {\bibfield
  {journal} {\bibinfo  {journal} {Phys. Rev. B}\ }\textbf {\bibinfo {volume}
  {73}},\ \bibinfo {pages} {035115} (\bibinfo {year} {2006})}\BibitemShut
  {NoStop}\bibitem [{\citenamefont {Beutel}\ \emph {et~al.}(2021)\citenamefont {Beutel},
  \citenamefont {Groner}, \citenamefont {Rockstuhl},\ and\ \citenamefont
  {Fernandez-Corbaton}}]{Beutel2020}\BibitemOpen
  \bibfield  {author} {\bibinfo {author} {\bibfnamefont {D.}~\bibnamefont
  {Beutel}}, \bibinfo {author} {\bibfnamefont {A.}~\bibnamefont {Groner}},
  \bibinfo {author} {\bibfnamefont {C.}~\bibnamefont {Rockstuhl}},\ and\
  \bibinfo {author} {\bibfnamefont {I.}~\bibnamefont {Fernandez-Corbaton}},\
  }\bibfield  {title} {\bibinfo {title} {Efficient simulation of biperiodic,
  layered structures based on the t-matrix method},\ }\href
  {https://doi.org/10.1364/JOSAB.419645} {\bibfield  {journal} {\bibinfo
  {journal} {J. Opt. Soc. Am. B}\ }\textbf {\bibinfo {volume} {38}},\ \bibinfo
  {pages} {1782} (\bibinfo {year} {2021})}\BibitemShut {NoStop}\bibitem [{\citenamefont {Vavilin}\ and\ \citenamefont
  {Fernandez-Corbaton}(2024)}]{Vavilin2023}\BibitemOpen
  \bibfield  {author} {\bibinfo {author} {\bibfnamefont {M.}~\bibnamefont
  {Vavilin}}\ and\ \bibinfo {author} {\bibfnamefont {I.}~\bibnamefont
  {Fernandez-Corbaton}},\ }\bibfield  {title} {\bibinfo {title} {The
  polychromatic t-matrix},\ }\href
  {https://doi.org/https://doi.org/10.1016/j.jqsrt.2023.108853} {\bibfield
  {journal} {\bibinfo  {journal} {JQSRT}\ }\textbf {\bibinfo {volume} {314}},\
  \bibinfo {pages} {108853} (\bibinfo {year} {2024})}\BibitemShut {NoStop}\bibitem [{\citenamefont {Gross}(1964)}]{Gross1964}\BibitemOpen
  \bibfield  {author} {\bibinfo {author} {\bibfnamefont {L.}~\bibnamefont
  {Gross}},\ }\bibfield  {title} {\bibinfo {title} {Norm invariance of
  {Mass‐Zero} equations under the conformal group},\ }\href
  {https://doi.org/10.1063/1.1704164} {\bibfield  {journal} {\bibinfo
  {journal} {J. Math. Phys.}\ }\textbf {\bibinfo {volume} {5}},\ \bibinfo
  {pages} {687} (\bibinfo {year} {1964})}\BibitemShut {NoStop}\bibitem [{\citenamefont {Zel'dovich}(1965)}]{Zeldovich1965}\BibitemOpen
  \bibfield  {author} {\bibinfo {author} {\bibfnamefont {Y.~B.}\ \bibnamefont
  {Zel'dovich}},\ }\bibfield  {title} {\bibinfo {title} {The number of quanta
  as an invariant of classical electromagnetic field},\ }\href@noop {}
  {\bibfield  {journal} {\bibinfo  {journal} {Doklady Akademii Nauk SSSR (USSR)
  English translation currently published in a number of subject-oriented
  journals}\ }\textbf {\bibinfo {volume} {163}} (\bibinfo {year}
  {1965})}\BibitemShut {NoStop}\bibitem [{\citenamefont {Ge}\ and\ \citenamefont {Hughes}(2015)}]{Ge2015}\BibitemOpen
  \bibfield  {author} {\bibinfo {author} {\bibfnamefont {R.-C.}\ \bibnamefont
  {Ge}}\ and\ \bibinfo {author} {\bibfnamefont {S.}~\bibnamefont {Hughes}},\
  }\bibfield  {title} {\bibinfo {title} {Quantum dynamics of two quantum dots
  coupled through localized plasmons: An intuitive and accurate quantum optics
  approach using quasinormal modes},\ }\href
  {https://doi.org/10.1103/PhysRevB.92.205420} {\bibfield  {journal} {\bibinfo
  {journal} {Phys. Rev. B}\ }\textbf {\bibinfo {volume} {92}},\ \bibinfo
  {pages} {205420} (\bibinfo {year} {2015})}\BibitemShut {NoStop}\bibitem [{\citenamefont {Waks}\ and\ \citenamefont
  {Sridharan}(2010)}]{Waks2010}\BibitemOpen
  \bibfield  {author} {\bibinfo {author} {\bibfnamefont {E.}~\bibnamefont
  {Waks}}\ and\ \bibinfo {author} {\bibfnamefont {D.}~\bibnamefont
  {Sridharan}},\ }\bibfield  {title} {\bibinfo {title} {Cavity qed treatment of
  interactions between a metal nanoparticle and a dipole emitter},\ }\href
  {https://doi.org/10.1103/PhysRevA.82.043845} {\bibfield  {journal} {\bibinfo
  {journal} {Phys. Rev. A}\ }\textbf {\bibinfo {volume} {82}},\ \bibinfo
  {pages} {043845} (\bibinfo {year} {2010})}\BibitemShut {NoStop}\bibitem [{\citenamefont {Zambrana-Puyalto}\ and\ \citenamefont
  {Bonod}(2016)}]{Zambrana2016b}\BibitemOpen
  \bibfield  {author} {\bibinfo {author} {\bibfnamefont {X.}~\bibnamefont
  {Zambrana-Puyalto}}\ and\ \bibinfo {author} {\bibfnamefont {N.}~\bibnamefont
  {Bonod}},\ }\bibfield  {title} {\bibinfo {title} {Tailoring the chirality of
  light emission with spherical si-based antennas},\ }\href
  {https://doi.org/10.1039/C6NR00676K} {\bibfield  {journal} {\bibinfo
  {journal} {Nanoscale.}\ }\textbf {\bibinfo {volume} {8}},\ \bibinfo {pages}
  {10441} (\bibinfo {year} {2016})}\BibitemShut {NoStop}\bibitem [{\citenamefont {Moses}(1965)}]{Moses1965b}\BibitemOpen
  \bibfield  {author} {\bibinfo {author} {\bibfnamefont {H.~E.}\ \bibnamefont
  {Moses}},\ }\bibfield  {title} {\bibinfo {title} {Transformation from a
  linear momentum to an angular momentum basis for relativistic particles of
  nonzero mass and any spin},\ }\href {https://doi.org/10.1063/1.1704766}
  {\bibfield  {journal} {\bibinfo  {journal} {Journal of Mathematical Physics}\
  }\textbf {\bibinfo {volume} {6}},\ \bibinfo {pages} {1244} (\bibinfo {year}
  {1965})}\BibitemShut {NoStop}\bibitem [{Sur()}]{SurfintCode2023}\BibitemOpen
  \href@noop {} {}\bibinfo {howpublished}
  {\url{https://www.waves.kit.edu/downloads/CRC1173_Preprint_2023_surfaces_Codes.zip}}\BibitemShut
  {NoStop}\bibitem [{\citenamefont {Bialynicki-Birula}(1996)}]{Birula1996}\BibitemOpen
  \bibfield  {author} {\bibinfo {author} {\bibfnamefont {I.}~\bibnamefont
  {Bialynicki-Birula}},\ }\bibfield  {title} {\bibinfo {title} {Photon wave
  function},\ }\href@noop {} {\bibfield  {journal} {\bibinfo  {journal} {Prog.
  Optics}\ }\textbf {\bibinfo {volume} {36}},\ \bibinfo {pages} {245} (\bibinfo
  {year} {1996})}\BibitemShut {NoStop}\bibitem [{\citenamefont {{Lakhtakia A.}}(1994)}]{Lakhtakia1994}\BibitemOpen
  \bibfield  {author} {\bibinfo {author} {\bibnamefont {{Lakhtakia A.}}},\
  }\href@noop {} {\emph {\bibinfo {title} {{Beltrami fields in chiral
  media}}}}\ (\bibinfo  {publisher} {World Scientific},\ \bibinfo {year}
  {1994})\BibitemShut {NoStop}\bibitem [{\citenamefont {Cohen-Tannoudji}\ \emph {et~al.}(1989)\citenamefont
  {Cohen-Tannoudji}, \citenamefont {Dupont-Roc},\ and\ \citenamefont
  {Grynberg}}]{Cohen1997}\BibitemOpen
  \bibfield  {author} {\bibinfo {author} {\bibfnamefont {C.}~\bibnamefont
  {Cohen-Tannoudji}}, \bibinfo {author} {\bibfnamefont {J.}~\bibnamefont
  {Dupont-Roc}},\ and\ \bibinfo {author} {\bibfnamefont {G.}~\bibnamefont
  {Grynberg}},\ }\href@noop {} {\emph {\bibinfo {title} {{Photons and Atoms:
  Introduction to Quantum Electrodynamics}}}}\ (\bibinfo  {publisher} {Wiley},\
  \bibinfo {year} {1989})\BibitemShut {NoStop}\bibitem [{\citenamefont {Bialynicki-Birula}\ and\ \citenamefont
  {Bialynicka-Birula}(1975)}]{Birula1975}\BibitemOpen
  \bibfield  {author} {\bibinfo {author} {\bibfnamefont {I.}~\bibnamefont
  {Bialynicki-Birula}}\ and\ \bibinfo {author} {\bibfnamefont {Z.}~\bibnamefont
  {Bialynicka-Birula}},\ }\href@noop {} {\emph {\bibinfo {title} {Quantum
  Electrodynamics}}}\ (\bibinfo  {publisher} {Pergamon, Oxford, UK},\ \bibinfo
  {year} {1975})\BibitemShut {NoStop}\bibitem [{\citenamefont {Fernandez-Corbaton}\ and\ \citenamefont
  {Vavilin}(2023)}]{FerCor2022b}\BibitemOpen
  \bibfield  {author} {\bibinfo {author} {\bibfnamefont {I.}~\bibnamefont
  {Fernandez-Corbaton}}\ and\ \bibinfo {author} {\bibfnamefont
  {M.}~\bibnamefont {Vavilin}},\ }\bibfield  {title} {\bibinfo {title} {A
  scalar product for computing fundamental quantities in matter},\ }\bibfield
  {journal} {\bibinfo  {journal} {Symmetry}\ }\textbf {\bibinfo {volume}
  {15}},\ \href {https://doi.org/10.3390/sym15101839} {10.3390/sym15101839}
  (\bibinfo {year} {2023})\BibitemShut {NoStop}\bibitem [{\citenamefont {Bateman}(1910)}]{Bateman1910}\BibitemOpen
  \bibfield  {author} {\bibinfo {author} {\bibfnamefont {H.}~\bibnamefont
  {Bateman}},\ }\bibfield  {title} {\bibinfo {title} {The transformation of the
  electrodynamical equations},\ }\href
  {https://doi.org/https://doi.org/10.1112/plms/s2-8.1.223} {\bibfield
  {journal} {\bibinfo  {journal} {Proceedings of the London Mathematical
  Society}\ }\textbf {\bibinfo {volume} {s2-8}},\ \bibinfo {pages} {223}
  (\bibinfo {year} {1910})}\BibitemShut {NoStop}\bibitem [{\citenamefont {Santiago}\ \emph {et~al.}(2019)\citenamefont
  {Santiago}, \citenamefont {Hammerschmidt}, \citenamefont {Burger},
  \citenamefont {Rockstuhl}, \citenamefont {Fernandez-Corbaton},\ and\
  \citenamefont {Zschiedrich}}]{Garcia2018}\BibitemOpen
  \bibfield  {author} {\bibinfo {author} {\bibfnamefont {X.~G.}\ \bibnamefont
  {Santiago}}, \bibinfo {author} {\bibfnamefont {M.}~\bibnamefont
  {Hammerschmidt}}, \bibinfo {author} {\bibfnamefont {S.}~\bibnamefont
  {Burger}}, \bibinfo {author} {\bibfnamefont {C.}~\bibnamefont {Rockstuhl}},
  \bibinfo {author} {\bibfnamefont {I.}~\bibnamefont {Fernandez-Corbaton}},\
  and\ \bibinfo {author} {\bibfnamefont {L.}~\bibnamefont {Zschiedrich}},\
  }\bibfield  {title} {\bibinfo {title} {Decomposition of scattered
  electromagnetic fields into vector spherical wave functions on surfaces with
  general shapes},\ }\href {https://doi.org/10.1103/PhysRevB.99.045406}
  {\bibfield  {journal} {\bibinfo  {journal} {Phys. Rev. B}\ }\textbf {\bibinfo
  {volume} {99}},\ \bibinfo {pages} {045406} (\bibinfo {year}
  {2019})}\BibitemShut {NoStop}\bibitem [{\citenamefont {Jackson}(1998)}]{Jackson1998}\BibitemOpen
  \bibfield  {author} {\bibinfo {author} {\bibfnamefont {J.~D.}\ \bibnamefont
  {Jackson}},\ }\href@noop {} {\emph {\bibinfo {title} {Classical
  Electrodynamics}}}\ (\bibinfo  {publisher} {Wiley},\ \bibinfo {address} {New
  York City},\ \bibinfo {year} {1998})\BibitemShut {NoStop}\bibitem [{\citenamefont {Tung}(1985)}]{Tung1985}\BibitemOpen
  \bibfield  {author} {\bibinfo {author} {\bibfnamefont {W.-K.}\ \bibnamefont
  {Tung}},\ }\href@noop {} {\emph {\bibinfo {title} {Group Theory in
  Physics}}}\ (\bibinfo  {publisher} {World Scientific: Singapore},\ \bibinfo
  {year} {1985})\BibitemShut {NoStop}\bibitem [{\citenamefont {Lasa-Alonso}\ \emph {et~al.}(2023)\citenamefont
  {Lasa-Alonso}, \citenamefont {Devescovi}, \citenamefont {Maciel-Escudero},
  \citenamefont {Garc{\'\i}a-Etxarri},\ and\ \citenamefont
  {Molina-Terriza}}]{Lasa2023}\BibitemOpen
  \bibfield  {author} {\bibinfo {author} {\bibfnamefont {J.}~\bibnamefont
  {Lasa-Alonso}}, \bibinfo {author} {\bibfnamefont {C.}~\bibnamefont
  {Devescovi}}, \bibinfo {author} {\bibfnamefont {C.}~\bibnamefont
  {Maciel-Escudero}}, \bibinfo {author} {\bibfnamefont {A.}~\bibnamefont
  {Garc{\'\i}a-Etxarri}},\ and\ \bibinfo {author} {\bibfnamefont
  {G.}~\bibnamefont {Molina-Terriza}},\ }\bibfield  {title} {\bibinfo {title}
  {On the origin of the kerker phenomena},\ }\href@noop {} {\bibfield
  {journal} {\bibinfo  {journal} {arXiv preprint arXiv:2306.12762}\ } (\bibinfo
  {year} {2023})}\BibitemShut {NoStop}\bibitem [{\citenamefont {Nieto-Vesperinas}(2017)}]{Vesperinas2017}\BibitemOpen
  \bibfield  {author} {\bibinfo {author} {\bibfnamefont {M.}~\bibnamefont
  {Nieto-Vesperinas}},\ }\bibfield  {title} {\bibinfo {title} {Chiral optical
  fields: a unified formulation of helicity scattered from particles and
  dichroism enhancement},\ }\href {https://doi.org/10.1098/rsta.2016.0314}
  {\bibfield  {journal} {\bibinfo  {journal} {Philosophical Transactions of the
  Royal Society A: Mathematical, Physical and Engineering Sciences}\ }\textbf
  {\bibinfo {volume} {375}},\ \bibinfo {pages} {20160314} (\bibinfo {year}
  {2017})}\BibitemShut {NoStop}\bibitem [{\citenamefont {Oskooi}\ \emph {et~al.}(2010)\citenamefont {Oskooi},
  \citenamefont {Roundy}, \citenamefont {Ibanescu}, \citenamefont {Bermel},
  \citenamefont {Joannopoulos},\ and\ \citenamefont {Johnson}}]{Oskooi2010}\BibitemOpen
  \bibfield  {author} {\bibinfo {author} {\bibfnamefont {A.~F.}\ \bibnamefont
  {Oskooi}}, \bibinfo {author} {\bibfnamefont {D.}~\bibnamefont {Roundy}},
  \bibinfo {author} {\bibfnamefont {M.}~\bibnamefont {Ibanescu}}, \bibinfo
  {author} {\bibfnamefont {P.}~\bibnamefont {Bermel}}, \bibinfo {author}
  {\bibfnamefont {J.}~\bibnamefont {Joannopoulos}},\ and\ \bibinfo {author}
  {\bibfnamefont {S.~G.}\ \bibnamefont {Johnson}},\ }\bibfield  {title}
  {\bibinfo {title} {Meep: A flexible free-software package for electromagnetic
  simulations by the fdtd method},\ }\href
  {https://doi.org/https://doi.org/10.1016/j.cpc.2009.11.008} {\bibfield
  {journal} {\bibinfo  {journal} {Computer Physics Communications}\ }\textbf
  {\bibinfo {volume} {181}},\ \bibinfo {pages} {687} (\bibinfo {year}
  {2010})}\BibitemShut {NoStop}\bibitem [{\citenamefont {Savasta}\ \emph {et~al.}(2002)\citenamefont
  {Savasta}, \citenamefont {Di~Stefano},\ and\ \citenamefont
  {Girlanda}}]{Savasta2002}\BibitemOpen
  \bibfield  {author} {\bibinfo {author} {\bibfnamefont {S.}~\bibnamefont
  {Savasta}}, \bibinfo {author} {\bibfnamefont {O.}~\bibnamefont
  {Di~Stefano}},\ and\ \bibinfo {author} {\bibfnamefont {R.}~\bibnamefont
  {Girlanda}},\ }\bibfield  {title} {\bibinfo {title} {Light quantization for
  arbitrary scattering systems},\ }\href
  {https://doi.org/10.1103/PhysRevA.65.043801} {\bibfield  {journal} {\bibinfo
  {journal} {Phys. Rev. A}\ }\textbf {\bibinfo {volume} {65}},\ \bibinfo
  {pages} {043801} (\bibinfo {year} {2002})}\BibitemShut {NoStop}\bibitem [{\citenamefont {Oppermann}\ \emph {et~al.}(2018)\citenamefont
  {Oppermann}, \citenamefont {Straubel}, \citenamefont {Fernandez-Corbaton},\
  and\ \citenamefont {Rockstuhl}}]{Oppermann2018}\BibitemOpen
  \bibfield  {author} {\bibinfo {author} {\bibfnamefont {J.}~\bibnamefont
  {Oppermann}}, \bibinfo {author} {\bibfnamefont {J.}~\bibnamefont {Straubel}},
  \bibinfo {author} {\bibfnamefont {I.}~\bibnamefont {Fernandez-Corbaton}},\
  and\ \bibinfo {author} {\bibfnamefont {C.}~\bibnamefont {Rockstuhl}},\
  }\bibfield  {title} {\bibinfo {title} {Normalization approach for scattering
  modes in classical and quantum electrodynamics},\ }\href
  {https://doi.org/10.1103/PhysRevA.97.052131} {\bibfield  {journal} {\bibinfo
  {journal} {Phys. Rev. A}\ }\textbf {\bibinfo {volume} {97}},\ \bibinfo
  {pages} {052131} (\bibinfo {year} {2018})}\BibitemShut {NoStop}\end{thebibliography}
\end{document}